# Oleylamine aging of PtNi nanoparticles giving enhanced functionality for the oxygen reduction reaction

Gerard M Leteba[1,6], Yi-Chi Wang[2,7,8], Thomas J A Slater[2,3], Rongsheng Cai[2], Conor Byrne[2], Christopher P Race[2], David R G Mitchell[4], Pieter B J Levecque,[1] Neil P Young[5], Alex Walton[2], Angus I Kirkland[3,5], Sarah J Haigh[2*], Candace I Lang[6*]

1. Catalysis Institute, Department of Chemical Engineering, University of Cape Town, Corner of Madiba Circle and South Lane, Rondebosch 7701, South Africa.
2. Department of Materials, University of Manchester, Manchester, M13 9PL, UK.
3. Electron Physical Sciences Imaging Centre, Diamond Light Source Ltd., Oxfordshire OX11 0DE, UK
4. Electron Microscopy Centre, Innovation Campus, University of Wollongong, Wollongong NSW 2517, Australia
5. Department of Materials, University of Oxford, Parks Road, Oxford, OX1 3PH, U.K.
6. School of Engineering, Macquarie University, NSW 2109 Australia
7. Beijing Institute of Nanoenergy and Nanosystems, Chinese Academy of Sciences, Beijing, 101400, China
8. School of Nanoscience and Technology, University of Chinese Academy of Sciences, Beijing, 100049, China

*sarah.haigh@manchester.ac.uk
*candace.lang@mq.edu.au

**ABSTRACT:** *We report a rapid solution-phase strategy to synthesize alloyed PtNi nanoparticles which demonstrate outstanding functionality for the oxygen reduction reaction (ORR). This one-pot co-reduction colloidal synthesis results in a monodisperse population of single-crystal nanoparticles of rhombic dodecahedral morphology, with Pt enriched edges and compositions close to $Pt_1Ni_2$. We use nanoscale 3D compositional analysis to reveal for the first time that oleylamine (OAm)-aging of the rhombic dodecahedral $Pt_1Ni_2$ particles results in Ni leaching from surface facets, producing aged particles with concave faceting, an exceptionally high surface area and a composition of $Pt_2Ni_1$. We show that the modified atomic nanostructures catalytically outperform the original PtNi rhombic dodecahedral particles by more than 2-fold and also yield improved cycling durability. Their functionality for the ORR far exceeds commercially available Pt/C nanoparticle electrocatalysts, both in terms of mass-specific activities (up to a 25-fold increase) and intrinsic area-specific activities (up to a 27-fold increase).*

Commercial carbon-supported platinum nanoparticles (Pt/C) are effective electrocatalysts for both the anodic hydrogen oxidation reaction (HOR, $4H_2 \rightarrow 4H^+ + 4e^-$) and the cathodic oxygen reduction reaction (ORR, $O_2 + 4H^+ + 4e^- \rightarrow 2H_2O$) in polymer electrolyte membrane fuel cells (PEMFCs).[1,2] The commercial viability of PEMFCs is, however, limited by the high cost of Pt and the sluggish kinetics of the ORR.[1,3,4] Research has shown that alloying Pt with 3d transition metals enhances catalytic functionality while reducing the Pt load, with potential cost savings.[2,3] Substantial efforts have therefore been directed at novel solution-phase synthesis methods for Pt-based nanoalloys, which can exhibit enhanced catalytic activity as a result of the surface structure and chemistry.[2,5,6,7] Li *et al*. have shown that de-alloyed Pt-Ni nanowires exhibit massively enhanced mass activity (52 times improvement versus commercial Pt/C),[8] which is attributed to the formation of ultrafine jagged nanowires with highly stressed, undercoordinated surface configurations. Up to now, this outperforms most of the Pt-Ni particle-based catalysts. Open-framework $Pt_3Ni$ nanostructures have been found to display enhanced functionality as a result of their high surface area, high-index surface facets and Pt enrichment at surface sites,[5] whereas introducing rougher or concave topography to Pt-Ni nanoparticles can provide a conservative improvement in ORR activities.[9,10] Crystal faceting is another important factor to optimise ORR catalytic activities. It is found that anisotropic morphologies bounded by {111} crystal planes, display enhanced catalytic performance compared to the lower-index ({100} and {110} bounded) surfaces due to the favourable adsorption of hydroxyl rather than oxygen on high-index surfaces.[11–13] However, all the available nanoparticle systems for ORR have limitations associated with the complexity of the synthesis route, their cycling stability or reduced catalytic function after prolonged operation.

The scalability and controllability of wet chemistry synthesis, makes it an attractive route for developing alloyed nanocatalysts with good control of surface morphology.[5,14,15] The nucleation and growth of metallic nanostructures from solution is governed by kinetic growth parameters as well as by the nature and strength of surfactants, local concentration, temperature etc.[15–17] Particularly for bimetallic systems, the shape, size, and composition of the product can be highly sensitive to the exact synthesis conditions, thus a robust synthetic route is highly desirable for subsequent scale up.[11,15,16].

Here, we report a facile, scalable and robust synthetic procedure for the preparation of monodisperse PtNi nanoalloys with controllable morphology and surface composition. This novel route uses tetrabutylammonium borohydride (TBAB) as the reductant; in a mixture of hydrophobic surfactants oleylamine (OAm) and octadecylamine (ODA), with oleic acid (OLEA) or trioctylamine (TOA) as the third surfactant component, and results in PtNi nanoparticles with a predominantly rhombic dodecahedral morphology. OAm is then added as a surfactant to deflocculate the aggregated nanoparticles and improve their processability by creating a homogeneous dispersion. We measured the ORR performance of our as-synthesised PtNi nanoparticles and the OAm-aged PtNi nanoparticles and found that both exhibit significantly enhanced ORR functionality and improved stability compared with commercially available Pt/C electrocatalysts, with ORR functionality values exceeding the DoE target.[8] We found that our OAm-aged PtNi nanoparticles catalytically outperformed the as-synthesized PtNi nanoparticles so used accurate 3D atomic model of the nanoparticles based on high-angle annular dark field scanning transmission electron microscopy (HAADF-STEM) single particle reconstruction and energy dispersive X-ray spectroscopy (EDXS) to understand the changes in morphology and elemental distribution due to aging. We found that OAm aging results in preferential etching to produce Pt rich, concave surfaces by leaching of Ni, which is likely responsible for the improved performance.

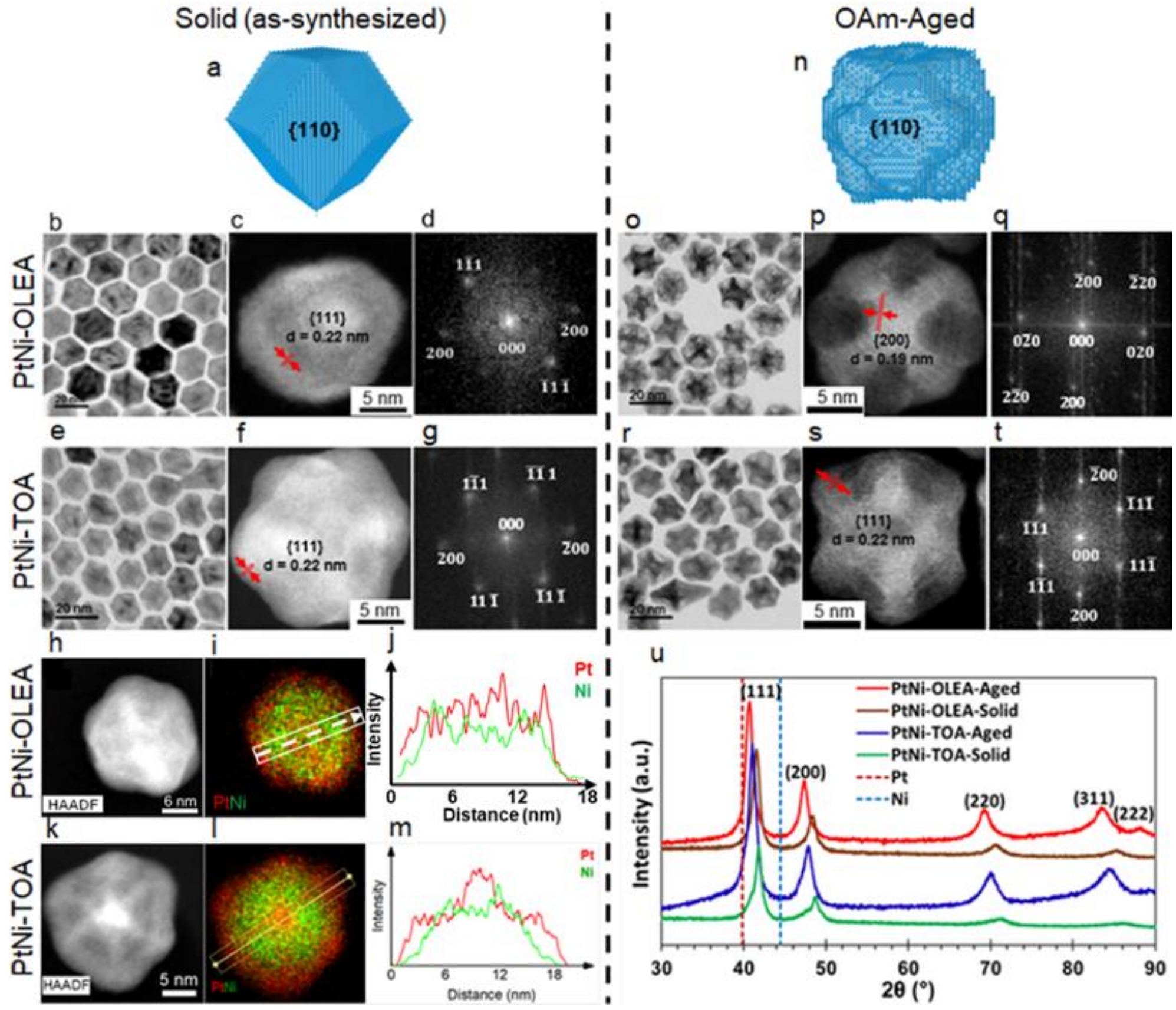

**Figure 1. Structural characterisation of PtNi nanoparticles.** (a-d) and (a, e-g) Characterisation of as-synthesised nanoparticle PtNi-OLEA-Solid and PtNi-TOA-Solid, respectively. (h-m) STEM-EDXS analysis of as-synthesized PtNi nanoparticles revealing the elemental distribution within single crystalline particles. (h, k) HAADF STEM images and (i, l) composite EDXS maps of Pt (red) and Ni (green) for PtNi-OLEA-Solid and PtNi-TOA-Solid, respectively. Line profiles extracted from the composite maps (i, l) are shown in (j, m) to illustrate the core-shell-shell NiPt nanoparticle structures. (n-q) and (n, r-t) Characterisation of aged nanoparticle PtNi-OLEA-Aged and PtNi-TOA-Aged, respectively. (a, n) Schematic structures of these nanoparticles before and after OAm-aging. (b, e, o, r) TEM images of nanoparticle populations. (c, f, p, s) High resolution HAADF-STEM images with measured lattice spacings and (d, g, q, t) are the corresponding fast Fourier transforms with indexed diffraction spots. (u) PXRD patterns of PtNi-OLEA and PtNi-TOA nanoparticles before and after aging showing fcc solid solutions of Pt and Ni. The positions of pure Pt(111) and Ni(111) peaks are indicated by dashed lines.

PtNi nanostructures were successfully solution-grown via a simultaneous reduction of nickel (II) acetate tetrahydrate and chloroplatinic acid solution (precursor salts), using TBAB as the reductant. The synthesis medium was a ternary mixture of hydrophobic surfactants OAm, ODA and OLEA in a high boiling point solvent 1-octadecene (1-OD). The effect of the surfactant OLEA was investigated by replacing OLEA with TOA, in this synthesis protocol. Particles synthesised with the OLEA surfactant are referred to here as PtNi-OLEA-Solid while those synthesised with TOA are referred to as PtNi-TOA-Solid. After aging in OAm the samples are referred to as PtNi-OLEA-Aged and PtNi-TOA-Aged for the OLEA synthesis and TOA synthesis respectively. We note here that nucleation was slow, followed by rapid growth. Our capacity to monitor the growth mechanisms of these colloidal alloys was accordingly limited.

The two as-synthesised types of PtNi alloy nanoparticles (PtNi-OLEA-Solid and PtNi-TOA-Solid) are observed to have similar morphologies, consisting of smoothly faceted rhombic dodecahedra with predominantly {110} facets (Figure 1(a-d) for PtNi-OLEA-Solid and Figure 1(a, e-g) for PtNi-TOA-Solid). For both PtNi-OLEA-Solid and PtNi-TOA-Solid, high quality monodispersed nanoparticles are observed. The uniformity in size and shape of the nanoparticles favours their self-arrangement/assembly into superlattice rafts when dispersed onto a thin carbon film suitable for transmission electron microscope (TEM) imaging.[18] Histograms showing the diameters of the as-synthesized nanoparticles, calculated from TEM images of 250–300 randomly selected individual nanoparticles, show a narrow average particle size distribution (means ± standard deviations) of PtNi-OLEA-Solid (17.3±1.6 nm), PtNi-TOA-Solid (16.8±1.3 nm), PtNi-OLEA-Aged (17.9±1.5 nm) and PtNi-TOA-Aged (16.6±1.5 nm) nanoparticles (Figure S1(a-d)), thus requiring no size-selective processing to achieve uniform particle size. Both TOA and OLEA protocols produced similar rhombic dodecahedral shaped nanoparticles, demonstrating that substituting the OLEA and TOA had little influence on the final synthesised structure. To further investigate the influence of reaction conditions the 1-OD solvent was substituted by an equal volume of benzyl ether (BE). This substitution also had no observable effect on the morphology of the nanoparticles produced (Figure S2). We conclude that the procedure is robust to changes in surfactants and solvents, highlighting the scale up potential of this nanostructure synthesis route.

We observe that the use of OAm in the initial co-reduction procedure results in excellent redispersion of the synthesized nanoparticles (required for detailed TEM analysis and for the homogeneous deposition on support materials) and also reverses the storage-instability of agglomerated nanoparticles. We further observe that this peptization phenomenon is only apparent in nanoparticles which have been solution-grown using hydrophobic surface coordinating surfactants. Unlike OAm, the use of organic nanoparticle stabilizers TOA and OLEA as the re-dispersants did not deflocculate the nanoparticle aggregates.

OAm also serves as an etchant to increase surface area. Aging as-synthesised nanoparticles in OAm for 3 weeks, we observe selective dissolution of nanoparticle surfaces during aging, creating concave facets in the rhombic dodecahedral structure although the overall size distribution is unaffected (Figure 1(n-t) and Figure S1).

Atomic structure of the Solid and OAm-Aged nanoparticles was examined by HAADF-STEM as shown in Figure 1(c, f, p, s). Atomic resolution images reveal the internal crystal structure and geometry of these nanoparticles. Fourier transforms (FTs) obtained from atomic resolution images of individual nanoparticles reveal these are perfect single crystals despite their unusual morphology (Figure 1(d, g, q, t)). Selected area electron diffraction (SAED) patterns acquired from ensembles of individual nanocrystals further demonstrate the single crystal nature of the nanostructures (Figure S1(e, f)), confirming the expected face centred cubic (fcc) lattice structure for all particles. A lattice parameter of 0.38±0.02 nm was measured for both PtNi-OLEA-Aged and PtNi-TOA-Aged samples via HAADF-STEM images (see Table S1).

Crystal structures of the PtNi-OLEA and PtNi-TOA nanoparticles before and after OAm-aging were further analysed by powder X-ray diffraction (PXRD) (Figure 1(u)) (calculated d-spacings and lattice parameters are shown in Table S1). Partial incorporation of smaller atoms like Ni into the Pt crystal lattice induces a shift to higher 2θ angles relative to pure Pt. This, in turn, facilitates ORR catalytic performance by creating a favourable surface platform to weaken oxygen binding energy or surface-adsorbed hydroxyl (OH) species on Pt.[19,20] The identified PXRD 2θ diffraction patterns are consistent with an fcc phase solid solution of Pt and Ni. A simple analysis of the {111} diffraction peaks using Vegard's rule gives the composition of the PtNi-OLEA-Solid and the PtNi-TOA-Solid particles as $Pt_{29}Ni_{71}$ and $Pt_{37}Ni_{63}$ respectively. The high angle diffraction peaks become more prominent after OAm-aging, suggesting lower lattice strain and enhanced nanoparticle crystallinity compared to the as synthesised structures. Aging of the PtNi nanoparticles is observed to result in a shift of the diffraction peaks towards lower angles: for example, the {111} diffraction peak of PtNi-TOA shifts from 2θ = 41.841° to 41.023°. This indicates that the interplanar spacings (d) are increased on aging from 0.216nm to 0.220nm, which is also in agreement with the change in d-spacing calculated from electron diffraction patterns (Table S1). The increasing d-spacing after aging is attributed to partial dissolution of Ni from the PtNi fcc lattice (Vegard analysis gives compositions of PtNi-OLEA-Aged and PtNi-TOA-Aged as $Pt_{54}Ni_{46}$ and $Pt_{49}Ni_{51}$).

Debye Scherrer analysis of the PXRD data revealed average nanoparticle crystallite sizes of 19.4±0.01 nm and 17.3±0.01 nm for PtNi-OLEA-Aged and PtNi-TOA-Aged respectively. Both these values are slightly larger than the average values obtained from the TEM/STEM size analysis although they are within the standard deviation of both techniques. It is not, however, unusual to find slightly higher average diameter measurements from XRD compared to TEM for single crystal nanostructures.[21] The STEM-EDXS spectrum imaging (Figure 1(h-m) and Figure S3) reveals elemental distributions within individual (PtNi-OLEA-Solid and the PtNi-TOA-Solid) nanoparticles at the nanometer scale. Intensity profiles acquired through the elemental maps reveal evidence of segregation within individual nanoparticles (Figures 1j and m) discussed in detail later.

To further characterise the unusual nanoparticle morphology after etching, the averaged 3D structure of the nanoparticle population was determined by HAADF-STEM single particle reconstruction using about 400 PtNi-OLEA-Aged nanoparticles (Figure 2 and Figure S4; for detail see experimental methods and ref[22]). The reconstruction revealed a 3D morphology consisting of a rhombic dodecahedron with concave facets. The rhombic dodecahedron morphology was further confirmed to be representative of the wider population by comparison of the projected 3D structure with experimental images for different nanocrystals (Figure S4). A full 3D atomic model (Figure 2(b and j)) was determined by combining the reconstructed mean 3D structure and crystallographic information from the atomic resolution TEM images (Figure 2(d and l)). Comparing multislice image simulations (Figures 2(c, e, f) and (k, m, n)) with experimental atomic resolution TEM images of aged nanoparticles with different crystal orientations (Figures 2(d, g, h) and (l, o, p)) reveals a good qualitative match between the broad intensity distribution and atomic structure.

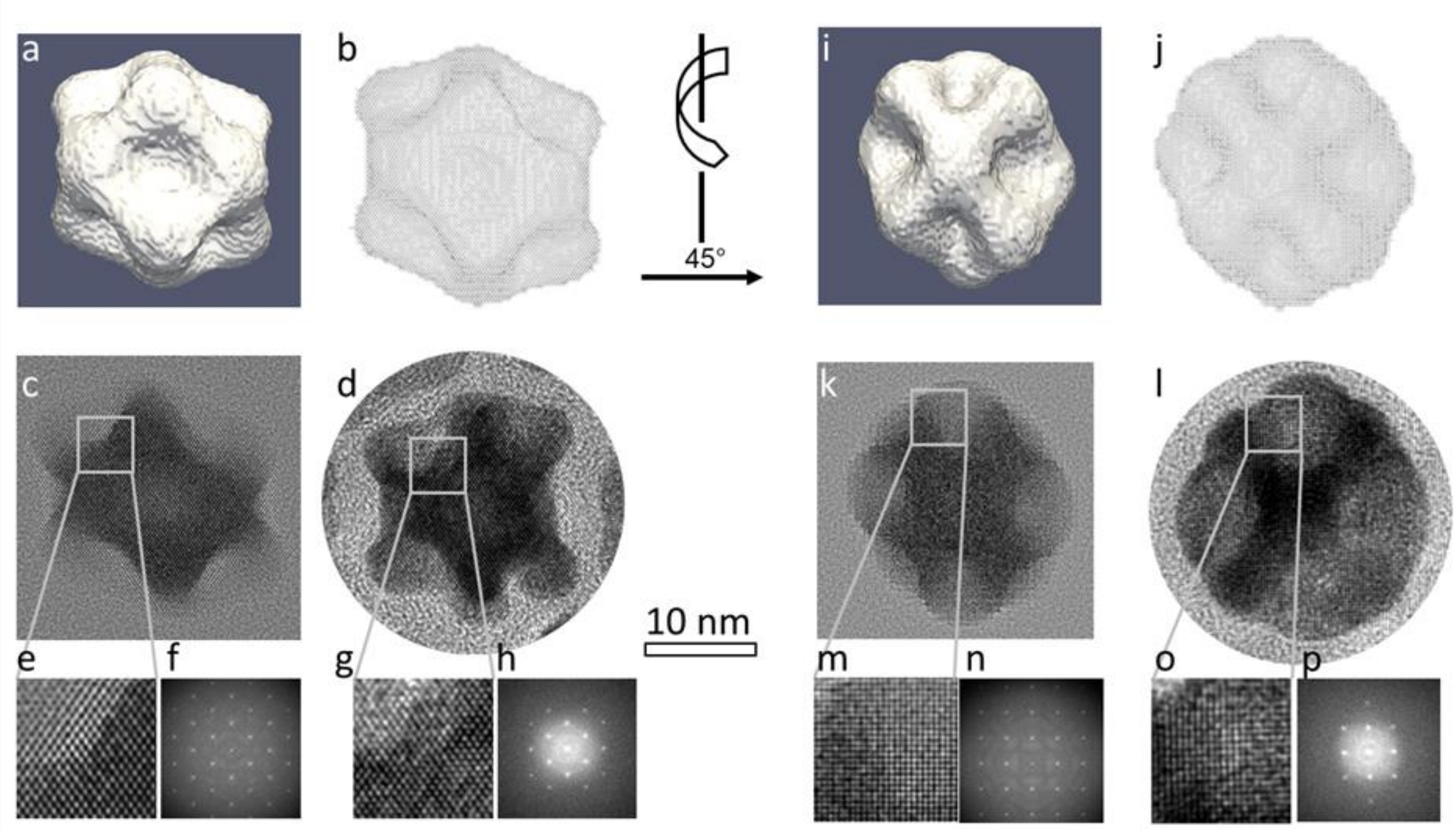

**Figure 2. Comparison of the 3D atomic reconstruction with experimental TEM images for PtNi-OLEA-Aged nanoparticles**, viewed along (a-h) a <110> zone-axis and (i-p) a <100> zone-axis. (a, i) Isosurface rendering of the 3D reconstruction representing the averaged morphology of PtNi-OLEA-Aged nanoparticles. (b, j) Visualization of the proposed atomic model of rhombic dodecahedral shape. (c, k) Simulated HRTEM images of the atomic model and (d, l) HRTEM images of an PtNi-OLEA-Aged nanoparticle. (e, m) Enlarged view of the HRTEM images for the atomic model in (c, k). (f, n) FFT of the HRTEM images of the atomic model. (g, o) Enlarged view of the HRTEM images in (d, l). (h, p) FFT of the real HRTEM images. Scale bars for (c, d, k, l) are 10 nm.

The composition and elemental distribution within these bimetallic nanoparticles before and after OAm-aging were then analysed by EDXS (Figure 3, Figure S3 and Figure S**5**). The average chemical composition of the nanoparticles was measured by summing the STEM-EDXS spectra obtained for individual particles and quantifying using a Cliff-Lorimer analysis (Figure S**6**). This showed the as-synthesized bimetallic nanocrystals to have mean compositions of $Pt_{32}Ni_{68}$ (PtNi-OLEA-Solid) and $Pt_{34}Ni_{66}$ (PtNi-TOA-Solid), close to $Pt_1Ni_2$ and consistent with the PXRD prediction. Subsequent to aging in OAm, the nanoparticles exhibited substantial reduction in their Ni content, with the nanoparticles having average compositions of $Pt_{64}Ni_{36}$ (PtNi-OLEA-Aged) and $Pt_{56}Ni_{44}$ (PtNi-TOA-Aged), close to $Pt_2Ni_1$ providing further evidence of OAm induced Ni surface and subsurface dissolution during aging.[5] Intensity profiles from the STEM-EDXS elemental maps reveal compositional inhomogeneity within individual nanoparticle structures (Figures 3(e-g) and (l-n)). The Pt-rich core could be the remnants of the Pt seed from nucleation at the initial co-reduction.[23] Note that this seed is usually close to the geometric centre of the particle but can also be found to be off-centre.

The preferential segregation of Pt to the apices or surfaces might be expected thermodynamically, as pure Pt has a larger lattice parameter than Ni so this segregation will reduce the total lattice strain of the system, consistent with the pure 'Pt skin' reported previously in model surfaces.[3,13,19] Based on atom column intensities and interatomic distances in atomic resolution HAADF images of an aged nanoparticle, the estimated Pt-rich skin is about 3-4 atom layers (Figure S7). The elemental distribution within PtNi-TOA nanoparticles revealed by STEM-EDXS was found to be similar (Figure S6). More detailed 3D characterisation via spectroscopic single particle reconstruction revealed that the aged PtNi nanoparticles possess concave Ni-rich surfaces and a Pt-rich outer frame, alongside the Pt-rich core.[22]

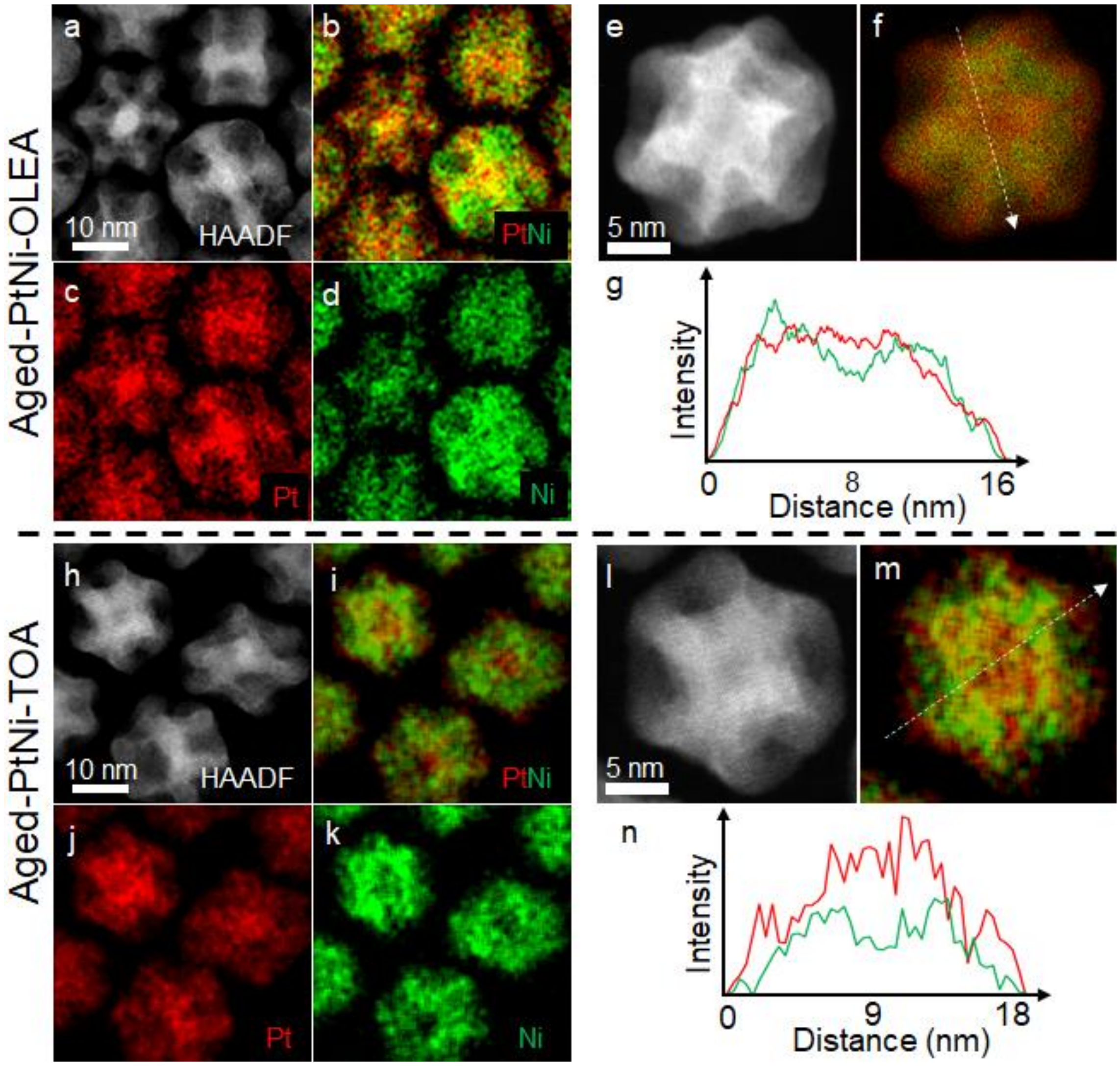

**Figure 3.** STEM-EDXS elemental distribution analysis of PtNi-OLEA (a-g) and PtNi-TOA (h-n) nanoparticles after OAm-aging. HAADF STEM images (a, e, h, l) are shown alongside STEM EDXS elemental maps for Pt (c, j) and Ni (d, k) in the same specimen region. Composite elemental maps demonstrate the relative locations of Pt and Ni (b, f, i, m). Line profiles extracted at the positions shown in the composite maps (f and m) are shown in (g and n) to illustrate the inhomogeneous spatial distributions of Pt and Ni within individual nanoparticle structures, as well as the Pt-enriched cores.

Further information on the effect of aging on composition was sought using X-ray photoelectron spectroscopy (XPS), as shown in Figure S8. In the as-prepared condition, PtNi nanoparticles exhibit a double peak in the Pt 4f region, corresponding to metallic Pt, which is also evident in aged samples. A small shift in position of the major Pt peak is observed after aging, reflecting a reduction in Ni content,[24] as also observed via TEM and PXRD after aging. Only a small peak corresponding to metallic Ni was detected (at around 852 eV) in as-prepared samples; this weak Ni peak is no longer evident in the C supported particles after aging.

The nanoparticle's monodispersity, high surface area and Pt-rich surface atop a PtNi subsurface is highly promising for their catalytic performance. To test the catalytic activity, four samples (PtNi-OLEA-Solid, PtNi-TOA-Solid, PtNi-OLEA-Aged and PtNi-TOA-Aged) were dispersed on highly conductive, high surface area carbon supports (Vulcan XC-72R), via a colloidal-deposition method (see Supporting Information methods for details). These carbon supported samples are refered to as PtNi-OLEA-Solid/C, PtNi-TOA-Solid/C, PtNi-OLEA-Aged/C and PtNi-TOA-Aged/C. Bright field (BF) TEM images of Solid-samples (Figures S9(a) and (b)) and Aged-samples (Figures 4(a) and (b)) show their good dispersion with no apparent alteration in surface structure and particle size distribution. This allowed electrochemical investigations of their true catalytic specific activities.

Both cyclic voltammograms (CV) (Figure 4(c)) and CO stripping curves (Figure S10(b), solid lines) were used to estimate the electrochemically active surface area (ECSA) of the synthesized nanocatalysts (20 wt.% loading) and the commercial Pt/C electrocatalyst (Alfa Aesar, HiSpec, 20 wt.% loading) (see Supplementary Information methods for details). The CV curves show the hydrogen adsorption/desorption (~0.05–0.35 V, *vs* standard hydrogen electrode, (SHE)) and oxide formation/reduction (~ 0.70–1.00 V, *vs* SHE) curves. Peak currents for these nanostructures were PtNi-TOA-Aged/C > PtNi-OLEA-Aged/C >> commercial Pt/C. Both $ECSA_{Hupd}$ ($_{Hupd}$ = under-potentially deposited hydrogen) and $ECSA_{CO}$ were calculated via normalization of the measured charges $Q_H$ (hydrogen adsorption) and $CO_{ads}$ (CO adsorption), adsorbed on the electrocatalysts using 210 μC/cm[25] and 420 μC/cm,[26] respectively. The $ECSA_{Hupd}$ scaled as commercial Pt/C (70.9 $m^2g_{Pt}^{-1}$) > PtNi-TOA-Aged (68.1 $m^2g_{Pt}^{-1}$) > PtNi-OLEA-Aged (65.7 $m^2g_{Pt}^{-1}$) whereas the $ECSA_{CO}$ scaled as: commercial Pt/C (87.2 $m^2g_{Pt}^{-1}$) > PtNi-TOA-Aged (74.7 $m^2g_{Pt}^{-1}$) > PtNi-OLEA-Aged/C (71.7 $m^2g_{Pt}^{-1}$) (Table S**2**).

The good ECSAs of these binary alloy nanoparticles are believed to be associated with the high concentration of surface defects such as hollow edges/corners associated with the complex 3D morphology demonstrated in Figure 2. In addition, electrochemical Ni dissolution (dealloying) during activation may have created nanoporous surfaces, further increasing the surface area of these nanostructures.[9,27,28] The $ECSA_{CO}/ECSA_{Hupd}$ ratios for these binary alloy nanoparticles were ~1.1, indicative of nominal differences in terms of both $H_{ads}$ and $CO_{ads}$ surface coverage. All the CO-stripping oxidation peaks for these nanostructures are located between +0.55 and +0.75 V (*vs* SHE). The presence of incorporated Ni just beneath the Pt surface resulted in the CO stripping peaks shifting to a lower potential than the commercial Pt/C electrocatalysts (between +0.6 and +0.85 V, *vs* SHE) (Figure 4(d)), suggesting that the presence of Ni incorporated into Pt surfaces improves CO tolerance through Pt electronic modification and weakening of the Pt-CO bond,[29,30] for both PtNi-OLEA-Aged/C and PtNi-TOA-Aged/C.

The ORR polarization curves show two regimes: (a) the mixed kinetic-diffusion controlled-region (the true measure of the catalyst functionality) in the region between +0.85 and +1.00 V (*vs* SHE) and (b) the diffusion limited current density regime from ~0.10 to ~0.85 V (*vs* SHE) (Figure 4(d)). All the polarization curves reached the diffusion limited-current density at ~ 6.0 mA $cm^{-2}$ (geometric) for ORR on all four PtNi nanostructures (Figure 4(d), consistent with reported theoretical values (~5.8–6.02 mA $cm^{-2}$).[4] This underlines the negligible influences of mass transport diffusion of molecular $O_2$ to the working electrode as a result of the homogeneity and thinness of the tested films. Inset Tafel plots (Figure 4(d) obtained from the potentials in the range of +0.85–0.95 V (*vs* SHE), exhibit the following activity trend: PtNi-OLEA-Aged/C > PtNi-TOA-Aged/C > PtNi-OLEA-Solid/C > PtNi-TOA-Solid/C >> commercial Pt/C, indicating the superior catalytic performance of these synthesised binary nanoparticles. The mass activities and specific activity at +0.90 V (*vs* SHE) were obtained by normalizing the kinetic currents ($I_k$) with the $ECSA_{Hupd}$ and the Pt catalyst mass immobilized on the electrode, respectively. The kinetic current ($I_k$) can be calculated using the Koutecky-Levich equation (Supplementary Information methods).[4]

The intrinsic area-specific activities (Figure 4(e)) of PtNi-OLEA-Aged/C and PtNi-TOA-Aged/C display ~27-times (1.39 mA $cm^{-2}$) and ~21-times (1.10 mA $cm^{-2}$) activity improvement, respectively, compared with the commercial Pt/C electrocatalyst (0.052 mA $cm^{-2}$). The Pt mass-specific activities of PtNi-OLEA-Aged/C and PtNi-TOA-Aged/C exhibit ~25-times (0.91 A/$mg_{Pt}$) and 20-times (0.75 A/$mg_{Pt}$) ORR enhanced functionality, respectively, compared with the commercial Pt/C electrocatalyst (0.037 A/$mg_{Pt}$) (Fig. 4(f)). We associate the higher activity of the PtNi-OLEA-Aged/C electrocatalyst with its more concave nanoparticle morphology, relative to the less concave PtNi-TOA-Aged/C electrocatalyst, resulting in greater surface area. The atomic surface sites on the concave surfaces of the PtNi-OLEA-Aged sample may also contribute to higher activity: previous studies have shown that electrocatalyst nanoparticles with stepped and terraced surfaces show improved catalytic activity when compared to nanoparticles with flat surfaces.[9,10]

The ORR area-specific activities of as-synthesized structures showed an increase of ~11 times (PtNi-OLEA-Solid/C, 0.59 mA $cm^{-2}$) and ~10 times (PtNi-TOA-Solid/C, 0.51 mA $cm^{-2}$) with respect to the commercial Pt/C electrocatalyst. The ORR mass-specific activities observed for PtNi-OLEA-Solid/C (0.33 A/$mg_{Pt}$) and PtNi-TOA-Solid/C (0.30 A/$mg_{Pt}$) nanostructures showed a slightly smaller increase: ~9-times and ~8-times higher than the commercial Pt/C, respectively. The OAm-aged nanostructures displayed a ~2-fold increase in terms of mass-specific activity and a ~3-fold enhancement of area-specific activities compared to the nanoalloys before aging. In addition, compared to the PtNi nanoframes by Chen et al.,[5] the PtNi-OLEA-Aged/C here (regarded as an intermediate morphology between nanoframes and the facetted rhombic dodecahedrons), shows a slightly greater improvement in mass activity (25-fold increase versus their 22-fold increase) and area specific activities (27-fold increase versus their 16-fold increase), where both are compared to the Pt/C references. These activity enhancements highlight the exigent need to exploit fine chemical surface etching techniques for the creation of next-generation advanced hybrid nanocatalysts for catalytic reactions, with uniquely uncoordinated Pt atomic surface arrangements. Table S3 compares the ORR activities for PtNi and other Pt alloys nanocatalysts from this and previous work. Although our system cannot currently compete with the best nanostructured PtNi catalysts: elaborately made with exotic morphologies consisting of jagged nanowires or nanoframes, the catalytic data presented here shows that these relatively solid PtNi particles (with ~2 times larger diameters) can achieve competitive specific activities compared with most PtNi nanoparticle catalysts by introduction of concave surfaces.

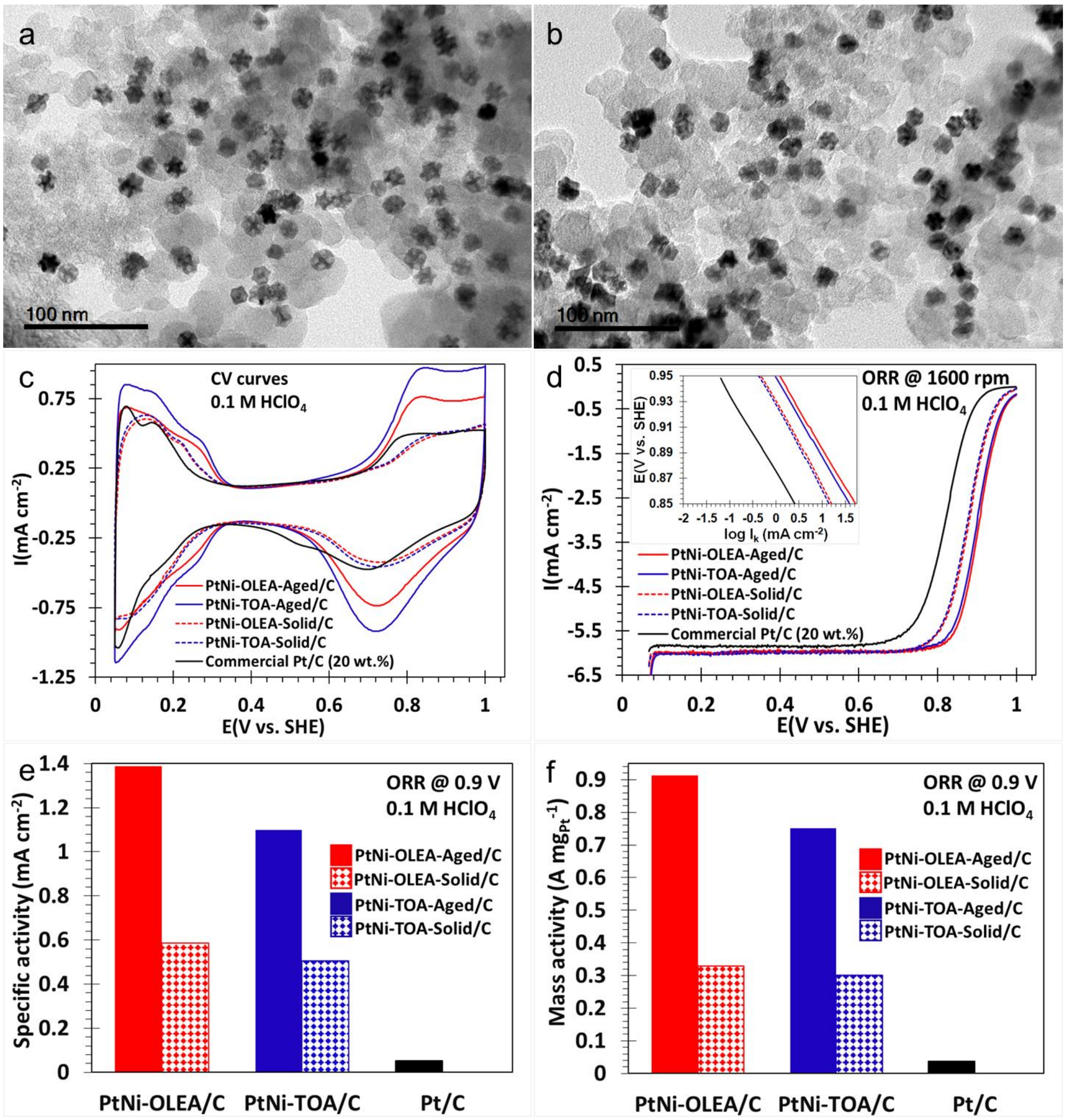

**Figure 4.** BF-TEM images of (a) PtNi-OLEA-Aged/C and (b) PtNi-TOA-Aged/C nanoparticles. (c) Cyclic voltammograms of activated binary PtNi-OLEA-Aged/C (red, solid plot), PtNi-OLEA-Solid/C (red, dashed plot) PtNi-TOA-Aged/C (blue, solid plot), PtNi-TOA-Solid/C (blue, dashed plot) and commercial Pt/C (black), (d) ORR polarization curves of samples before and after OAm aging (insert is the corresponding Tafel plots), (e) intrinsic area-specific activities and (f) mass-specific activities at +0.90 V (vs SHE) after 30 potential cycles of activation of electrocatalysts.

The development of durable electrocatalysts has been the focus of much recent research.[5,31–33] In the present work, the durability tests were conducted by subjecting the working electrode of the Aged-electrocatalysts to potential cycling between 0.05 V and 1.00 V at a scan rate of 100 mV/s for 5000 cycles in an Ar-purged 0.1 M $HClO_4$ electrolyte solution. This allowed the investigation of the electrocatalysts' ORR performance in addition to microstructural changes. CV measurements after 5000 electrochemical cycles of long-term in-service tests show apparent decrease in both the hydrogen desorption/adsorption and oxide formation/reduction peaks. The evolution of two prominent hydrogen adsorption ($H_{ads}$) current peaks, ascribed to {100} and {110} facets, corresponds to that of pure polycrystalline Pt[34] (Figure S10). Thus, the $ECSAs_{Hupd}$ of our nanoalloys decreased, relative to their initial value, by 40% for PtNi-OLEA-Aged/C and 36% for PtNi-TOA-Aged/C (Figure S11 and Table S2). In addition, recorded CO stripping curves display significant drop in current peaks ($ECSA_{CO}$ loss of 40% for PtNi-OLEA-Aged/C and 33% for PtNi-TOA-Aged/C) and positive potential shift (from lower to higher) (Figure S10 and Table S2). All these observations suggest selective Ni atomic surface and near-surface dissolution from the bulk alloys, resulting in diminished electrocatalytic performance and CO tolerance of the dealloyed nanoparticles. These binary electrocatalysts also exhibited substantial ORR functionality deterioration (onset potential shifting to lower potentials) and consequently the corresponding Tafel plots (Figure S10). These deteriorations in $ECSA_{Hupd}$, $ECSA_{CO}$ and overall catalytic activities of these binary nanostructures with respect to prolonged potential cycling could arise from a number of factors including electrochemical Ni dissolution, particle surface migration on the carbon support followed by coalescence/Ostwald ripening growth mechanism, or metal alloy oxide or hydroxide formation as a result of prolonged potential cycling and morphological deformations.[31,35] STEM-EDXS analysis revealed substantial changes in the nanoparticle size, morphological deformation and atomic rearrangements with particles becoming more spherical although Pt-enrichment is still apparent on the particle surface after cycling (Figure S12). In addition, more metal oxides exposed on the electrocatalysts surface may suppress the adsorption or diffusion of molecular $O_2$ species to the working electrode interface and thus hamper ORR performance.

XPS spectra (Figure S8) showed that the metallic Ni peak, visible in as-prepared nanoparticles, was absent in the C-supported nanoparticles after aging and after durability testing. The Pt peak remained well defined on the supported materials even after durability testing. A small Pt peak shift (from 71.0 eV to 71.5 eV) was observed after durability testing, which may suggest a decrease in Pt at the particle surface[36] although this was not detectable by STEM-EDXS.

To investigate the possible influence of surfactants on ECSA, we conducted Fourier transform infrared (FT-IR) measurements on all samples, as shown in Figure S13. The FT-IR spectra are in good agreement with previous reports.[37,38] The spectra of as-synthesized PtNi and aged PtNi nanoparticles shown in Figure S13 are similar to that of OAm, but not to OLEA, ODA or TOA. These FT-IR spectral investigations thus suggest that the organic layers passivating the surfaces of all nanoparticles (after repeated destabilization/purification processes) are only OAm. We would therefore expect only OAm to have a possible influence on the ECSA measurements. However, the higher ECSA values and significant positive shifts in the ORR polarization curves, in addition to the well-defined diffusion limited current density plateaux (reaching the theoretical value of ~6 mA $cm^{-2}$) suggest minimal interference from OAm during the catalytic performance of these PtNi systems. We therefore conclude that the influence of surfactants on catalytic activity is not significant. We also observe shifts in the diffusion limited current density regime from ~6.0 to ~5.8 mA/$cm^2$ post 5000 electrochemical cycles, suggesting restricted mass transport ($O_2$) diffusion and this may be induced by these electrocatalysts microstructural deformations or blockage/deactivation of Pt active sites due to surface contamination (Figure S10).

The results suggest that ORR functionality of the surface-aged PtNi alloy nanostructures is consistent with these newly evolved crystal facets via preferential OAm-Ni surface detachment. This is predicted to be further enhanced by electrochemical dealloying during the activation mechanism,[39] thus altering the surface electronic properties and creating novel Ni subsurface distribution/rearrangement within the Pt nanostructures. In addition, surface defects such as hollows/corners, interfaces, high density of atomic steps and kinks, as well as distinct crystallographic planes, are believed to offer improved accessibility of reacting molecules to the catalyst surface active sites.[5,10,27,40]

In conclusion, the facile and robust solution-based synthetic approach reported here resulted in the formation of highly monodisperse, crystalline PtNi alloyed nanoparticles with composition ~$Pt_1Ni_2$. Aging in the surfactant OAm, stabilised the structure and aged the surface of the nanoparticles, selectively dissolving Ni from the faces of the particles. This resulted in formation of concave surfaces and many Pt-rich apices, giving a mean nanoalloy composition close to $Pt_2Ni_1$. The surface-aged rhombic dodecahedral nanoparticles exhibited excellent electrocatalytic activity in the ORR, with PtNi mass-specific activities up to 25 times greater than a commercial Pt/C electrocatalyst. Preliminary durability measurements showed that mass-specific activity decayed, but only to a value which is still double the initial value for a commercial Pt/C electrocatalyst.

**Acknowledgements**

SJH and YW acknowledge funding from the European Research Council (ERC) under the European Union's Horizon 2020 research and innovation program under grant agreement No 715502 (EvoluTEM), the Chinese Scholarship Council and the EPSRC (UK) grant number EP/P009050/1. GML, PBJL and CIL acknowledge funding from the Hydrogen South Africa (HySA) and Macquarie University scholarship. DRGM acknowledges funding from the Australian Research Council (ARC), Linkage, Infrastructure, Equipment and Facilities (LIEF) grant number LE120100104. TEM and XPS access was supported by the Henry Royce Institute for Advanced Materials, funded through EPSRC grants EP/R00661X/1, EP/S019367/1, EP/P025021/1 and EP/P025498/1.

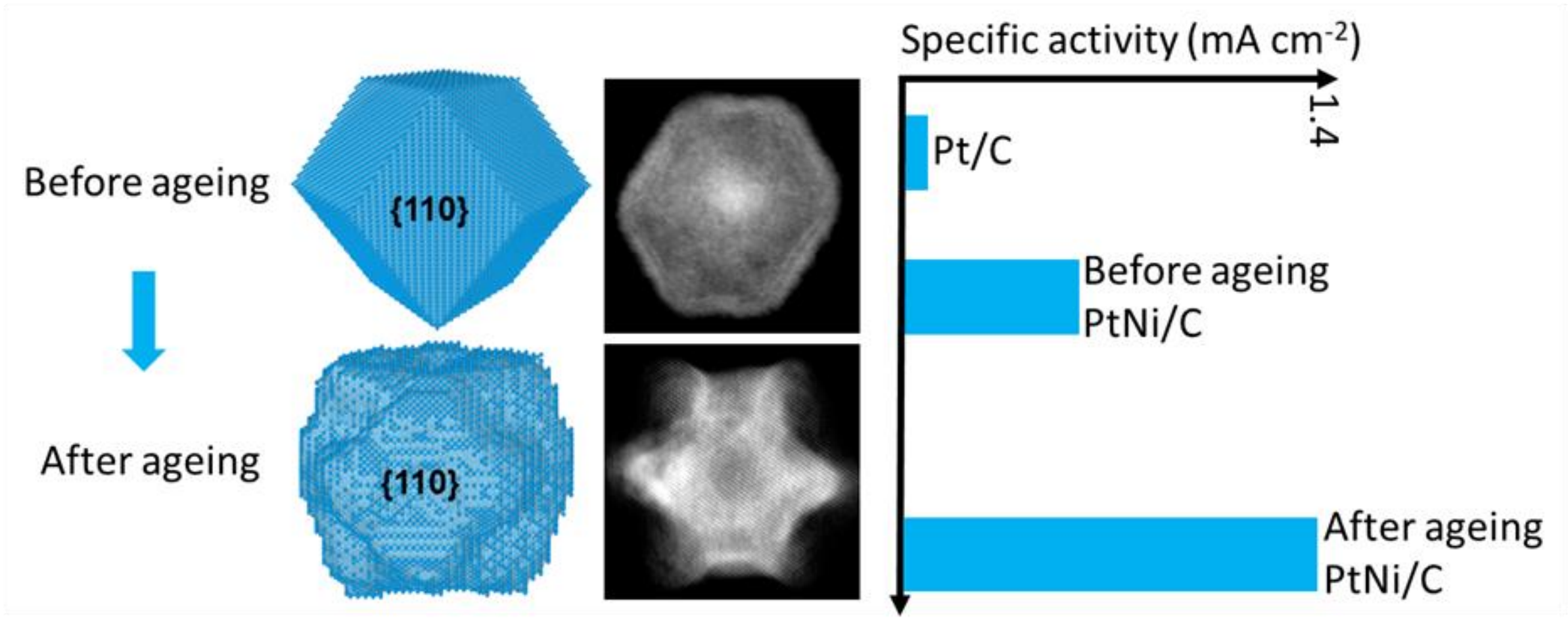

Table of Contents Graphic

# Supporting Information for

# "Oleylamine aging of PtNi nanoparticles giving enhanced functionality for the oxygen reduction reaction"

Gerard M Leteba[1,6], Yi-Chi Wang[2,7,8], Thomas J A Slater[2,3], Rongsheng Cai[2], Conor Byrne[2], Christopher P Race[2], David R G Mitchell[4], Pieter B J Levecque,[1] Alex Walton[2], Neil P Young[5], Angus I Kirkland[3,5], Sarah J Haigh[2*], Candace I Lang[6*]

1. Catalysis Institute, Department of Chemical Engineering, University of Cape Town, Corner of Madiba Circle and South Lane, Rondebosch 7701, South Africa.
2. Department of Materials, University of Manchester, Manchester, M13 9PL, UK.
3. Electron Physical Sciences Imaging Centre, Diamond Light Source Ltd., Oxfordshire OX11 0DE, UK
4. Electron Microscopy Centre, Innovation Campus, University of Wollongong, Wollongong NSW 2517, Australia
5. Department of Materials, University of Oxford, Parks Road, Oxford, OX1 3PH, U.K.
6. School of Engineering, Macquarie University, NSW 2109 Australia
7. Beijing Institute of Nanoenergy and Nanosystems, Chinese Academy of Sciences, Beijing, 101400, China
8. School of Nanoscience and Technology, University of Chinese Academy of Sciences, Beijing, 100049, China

*sarah.haigh@manchester.ac.uk
*candace.lang@mq.edu.au

## Experimental Methods:

### Synthesis of nanoparticles.

In a standard co-reduction procedure: 0.03 g nickel (II) acetate tetrahydrate ($Ni(Ac)_2 \cdot 4H_2O$) and 0.09 g chloroplatinic acid solution ($H_2PtCl_6$, 8 wt. % in water) (precursor salts), 15 ml oleylamine (OAm), 4.4 g octadecylamine (ODA) and 15 ml oleic acid (OLEA) (hydrophobic surfactants) were dissolved in 25 ml 1-octadecene (1-OD) (a high boiling point solvent) by sonication for 20 minutes; then heated at 150 °C under vigorous magnetic stirring until a pale yellow solution was observed. After the addition of 0.05 g tetrabutylammonium borohydride (TBAB) as the reductant, the reaction temperature was raised to 240 °C and maintained for 30–40 minutes in air. (The effect of the surfactant OLEA was investigated by repeating this protocol, replacing OLEA with TOA). The nanoparticles were aged in OAm for 3 weeks , followed by the addition of excess chloroform, purification by the addition of ethanol and finally, re-dispersing the particles in chloroform by vigorous sonication. The separation-precipitation washing process was performed three times to eliminate any unbound surfactants on the surfaces of the nanoparticles Thereafter the black product was dried and finally re-suspended in chloroform.

### Deposition of nanomaterials onto carbon support.

The as-prepared nanostructures were dispersed onto a carbon support (Cabot, Vulcan XC-72R) via a colloidal-deposition strategy, by mixing these nanomaterials and chloroform, followed by sonication for 15–20 min. The resulting homogeneous reaction dispersion was left in a fume hood overnight to evaporate the chloroform. The resultant carbon-supported materials were further washed with acetone 3–4 times and dried in an oven at 60 ºC. The metal loading (wt. % metal) for each sample was verified using thermogravimetric analysis (Mettler Toledo TGA/sDTA851e).

**Nanostructure characterization techniques.** The black powders of the as-synthesised nanostructures were deposited onto a zero-background silicon (Si) wafer support and characterized by powder XRD on an X'Pert Pro multipurpose diffractometer (MPD), using Cu $K_\alpha$ radiation ($\lambda = 0.154056$ nm). The diffraction patterns were recorded at a scan rate of 0.106º/s and with a step size of 0.0334º. Specimens for scanning transmission electron microscopy (STEM) investigations were prepared by drop-casting a colloidal solution onto 3-mm carbon-supported films on copper grids. These were air dried under ambient conditions. STEM-HAADF and EDS data were collected using a Thermo Fisher Titan G2 80-200 S/TEM operated at 200 kV, which was equipped with an X-FEG high brightness source, STEM probe aberration correction and a ChemiSTEM$^{TM}$ Super-X EDS detector consisting of four silicon drift detectors (SDDs) with a total collection solid angle of approximately 0.7 sr. A convergence angle of 21 mrad and an acceptance inner angle of 55 mrad were used for HAADF STEM image acquisition. Tilt-series tomography HAADF images were collected using a Fischione 2020 single tilt tomography holder and FEI Inspect3D software. The total tilt range was ±70° with a pixel size of 0.07 nm and a pixel dwell time of 10 µs. Incremental steps of 2° at ±50° and 1° for the rest of the tilt range were used. Nanoparticle size refers to the particle's diameter. To measure this from an HRTEM image of an individual particle two orthogonal lines were drawn across the particle, the first being aligned with the particle's largest dimension. The diameter was then determined as the average of the two line lengths.

The 3D reconstruction shown in Figure 2 was performed using a single particle reconstruction method, averaging approximately 400 PtNi-OLEA-Aged nanoparticles. As these nanoparticles are similar in shape and randomly oriented on the support, several STEM-HAADF images of different areas can collect data for about 1000 individual nanoparticles at various orientations. Nanoparticle orientations were solved by cross-correlating the experimental images with the reprojections from a conventional tilt series tomogram. Fourier-space back projection was used for 3D reconstruction after the orientation of each HAADF image was assigned. Detailed methods for this single particle reconstruction technique of these inorganic nanoparticles are described elsewhere.[1] Simulation of HRTEM images was performed using QSTEM[2] with the atomic model as an input with parameters of 200 kV accelarating voltage, 61.3 nm defocus, 0.5 mrad convergence angle, 3 nm focal spread, 0 nm astigmatism, 1 mm 3rd order speherical aberration and amorphous noise. To measure the local interatomic spacings from HAADF STEM data, each atom column position was identified using 2D Gaussian fitting in the python package atomap.[3] For each atom column, the distances to its surrounding atoms were calculated and averaged. This distance was then used to colour the atomic columns.

Specimens for X-ray photoelectron spectroscopy (XPS) were prepared by dropcasting suspensions (dispersed in chloroform) onto silicon wafers. High Resolution XPS Spectra were taken using a SPECS NAP-XPS instrument (operating under UHV conditions, base pressure of $1\times10^{-9}$ mbar). XPS measurements were taken using a microfocussed, monochromated Al $K_{\alpha}$ X – Ray source (1487eV) and a SPECS Phoibos NAP 150 hemispherical analyser. Scans were taken at normal emission and a pass energy of 30 eV for detailed scans and 60 eV for survey scans. The samples were charge referenced to adventitious carbon at 284.8 eV.

**Deposition of nanomaterials onto carbon support.** The as-prepared nanostructures were dispersed onto a carbon support (Cabot, Vulcan XC-72R) via a colloidal-deposition strategy, by mixing these nanomaterials and chloroform, followed by sonication for 15 min. The resulting homogeneous reaction dispersion was left in a fume hood for 5 hours to evaporate the chloroform. The resultant carbon-supported materials were further washed with acetone 5 times and dried in an oven at 60 ºC. The metal loading (wt.% metal) for each sample was verified using thermogravimetric analysis (Mettler Toledo TGA/sDTA851e).

**Electrochemical Measurements.** All experiments were performed in a standard three electrode setup at room temperature in a 0.1 M perchloric acid ($HClO_4$, 70%) solution using either argon (Ar) 99.999% (Afrox), oxygen ($O_2$) 99.998% (Afrox) and carbon dioxide (CO) 99% (Afrox) as specified. A platinum (Pt) coil was used as the counter electrode. A mercury/mercurous sulphate reference electrode was used as a reference and all potential values were reported relative to the standard hydrogen electrode (SHE). The readout currents were not corrected for the ohmic iR losses. A Biologic SP300 potentiostat was coupled to a RDE710 Rotator (Gamry instrument). Catalyst inks were prepared by mixing 10 mg of the catalyst with 2 ml of Milli-Q water (Millipore, 18 MΩ.cm @ 25 °C), 0.45 ml isopropanol and 25 µl Nafion® perfluorinated resin solution (5 wt.% in a mixture of lower aliphatic alcohols and water (45% water)). The resultant mixture was sonicated for 15–20 min. 10 µl of the ink was then pipetted onto a glassy carbon (GC) electrode (Pine Research Instrumentation, 5 mm disk OD) and dried under ambient conditions for 30–60 minutes to evaporate the solvents. The remaining thin black uniform film of Nafion-catalyst-Vulcan on the GC served as the working electrode (WE). Before use the GC electrode was polished to a mirror finish on a Microcloth polishing pad (Buehler) using 1 µm and 0.05 µm alumina paste (Buehler). After rinsing, the WE was ultrasonicated in Milli-Q water (Millipore, 18 MΩ.cm @ 25 °C) for 10 min and left to dry.

**(a) Cyclic voltammetry.**

In an Ar purged electrolyte, the potential of the WE was cycled between 0.05 V and 1.00 V vs. SHE at a scan rate of 100 mV/s for 30 cycles to electrochemically clean the catalyst surface. The sweep rate was then reduced to 50 mV/s and the third cycle at that scan rate was used for analysis. The electrochemically active surface area (ECSA) was calculated by integrating the area under the curve for the hydrogen underpotential deposition region ($H_{upd}$) assuming a monolayer hydrogen charge of 210 $\mu C/cm^2_{Pt}$. [4-5]

**(b) Carbon dioxide (CO) stripping voltammetry.**

CO gas was bubbled into the electrolyte solution while holding the potential of the working electrode at 0.1 V vs. SHE. The electrolyte was then purged with Ar to remove the dissolved CO gas while still holding the potential of the WE at 0.1 V vs. SHE. The potential of the WE was then cycled to 1.00 V vs. SHE at 20 mV/s, followed by a CV cycle as described above at 20 mV/s. The peak area could then be determined using the baseline CV and a normalisation factor of 420 $\mu g/cm^2_{Pt}$[6] was used to calculate the ECSA.

**(c) Linear Sweep Voltammetry.**

The potential of the WE was swept from 1.10 V to 0.20 V vs. SHE and back at 10 mV/s. ORR polarization curves were recorded at rotation speeds of 400, 900, 1600 and 2500 rpm. The ORR curves obtained in $O_2$ saturated electrolyte were corrected for the capacitive current associated with $Pt_xM_y/C$ catalysts, by subtracting a CV curve measured in an argon-saturated electrolyte. All the polarization curves reported in this work were acquired using a rotation speed of 1600 rpm and the current densities were also normalized with reference to the calculated ECSA to evaluate the specific activities.

The current density ($i$) for the ORR electrocatalytic activity is calculated according to Koutecky-Levich equation:[7-8]

$$\frac{1}{i} = \frac{1}{i_k} + \frac{1}{i_d}$$

where $i$ is the overall disk current density, $i_k$ is the true kinetic current density ($A$) and is determined by the mass transport properties of the RDE, $i_d$ is the diffusion limited current density. $i_d$ can be expressed according to Levich equation as follows:

$$i_d = 0.201 n_e F A D_{O_2}{}^{2/3} V^{-1/6} C_{O_2}{}^{\infty} \omega^{1/2}$$

where $n_e$ is the total number of electrons transferred ($4e^-$), $F$ is the Faraday's constant (96485 C/mol), $A$ is the surface area of the electrode (0.196 cm$^2$), $D_{O_2}$is the diffusion coefficient of oxygen ($1.93 \times 10^{-5}$ cm$^2$/s), $C_{O_2}$ is the concentration of dissolved oxygen ($1.26 \times 10^{-6}$ mol/cm$^3$), $V$ is the kinematic viscosity of the electrolyte solution ($1.01 \times 10^{-2}$ cm$^2$/s) at 20 °C and $\omega$ is the angular frequency of rotation, $\omega = \frac{2\pi f}{60}$, $f$ is the RDE rotation rate in rpm: for the measurements conducted in 0.1 M $HClO_4$ at 20 °C and 1 atm $O_2$. The diffusion coefficient of oxygen, the kinematic viscosity of the electrolyte solution and the concentration of dissolved oxygen are classified as non-electrochemical kinetic characteristics required for RDE data analysis. These kinetic parameters are influenced by temperature and the electrolyte solution during the electrochemical measurements.[9]

In order to determine the mass-transport free kinetic current ($i_k$), the ORR measurements are conducted at the rotation speed of 1600 rpm. This rotation rate is used as the benchmark to compare the functionality of electrocatalysts and the ORR polarization limiting current ranges between $5.8 \times 10^{-3}$ – $6 \times 10^{-3}$ mA cm$^{-2}$.[7, 10] These limiting current values yield n = 4 for the Levich and Levich-Koutecky plots. Furthermore, background current measurements are performed in deaerated electrolyte solution to account for capacitive current interferences. The difference between the experimentally measured current and the background current yields mass-transport corrected current. This current is used to evaluate mass-and area-specific activities of catalysts. Since at the limiting current the reaction kinetics occur very fast, the Koutecky-Levich equation can be re-arranged as follows:[8, 10-11]

$$I_k\,(A) = \frac{I_{lim}\,(A) \times I(A)}{(I_{lim} - I)\,(A)}$$

where $I_{lim} = i_d$ is the measured diffusion limited current density and $I_k$ is the kinetic current ($A$). The $I$ and $I_{lim}$ are the values calculated from the anodic ORR polarization curve at $E$ = 0.9 V and $E$ = 0.4 V versus SHE, respectively.[11]

The Pt mass-specific ($I_m$) and area-specific ($I_s$) activities are quantified at $E$ = 0.9 V versus SHE specifically because the contributions from mass-transport losses cannot be totally disregarded at the higher current densities detected below $E$ = 0.9 V.[9-12] Therefore, the Pt mass-specific activity is calculated from the $I_k$ and normalization to the Pt-loading of the GC disk electrode:[9-11]

$$I_{m(0.90V)}(A\ \mathrm{mg_{Pt}}^{-1}) = \frac{I_k\,(A)}{L_{\mathrm{Pt}}(\mathrm{mg_{Pt}\,cm^{-2}})(\mathrm{Ag\,cm^{2}})}$$

$L_{Pt}$ is the working electrode catalyst loading ($mg_{catalyst}$ $cm^{-2}$) and $Ag$ ($cm^2$) is the geometric surface area of the glassy carbon electrode. The area-specific activity is determined from the $I_k$ and normalization with the Pt electrochemical surface area (ECSA):[10-11, 13]

$$I_{s(0.90V)}(mA/cm^2) = \frac{I_k\ (A)}{ECSA}$$

$$= \frac{I_k\ (A)}{(Q_{H-\text{desorption (or adsorption)}}(\text{C})/210\ \mu\text{C cm}^{-2})Ag(\text{cm}^2)}$$

In order to investigate the possible influence of surfactants on ECSA, we conducted Fourier transform infrared (FT-IR) measurements on all samples, as shown in Figure S13. The FT-IR spectra are in good agreement with previous reports.[14-15] The spectra of as-synthesized PtNi and aged PtNi nanoparticles shown in Figure S11 are similar to that of OAm, but not to OLEA, ODA or TOA. These FT-IR spectral investigations thus suggest that the organic layers passivating the surfaces of all nanoparticles (after repeated destabilization/purification processes) are only OAm. We would therefore expect only OAm to have a possible influence on the ECSA measurements. However, the higher ECSA values and significant positive shifts in the ORR polarization curves, in addition to the well-defined diffusion limited current density plateaux (reaching the theoretical value of ~6 $mA/cm^2$) suggest minimal interference of OAm during the catalytic performance of these PtNi systems. We accordingly concluded that the influence of surfactants on catalytic activity is not significant.

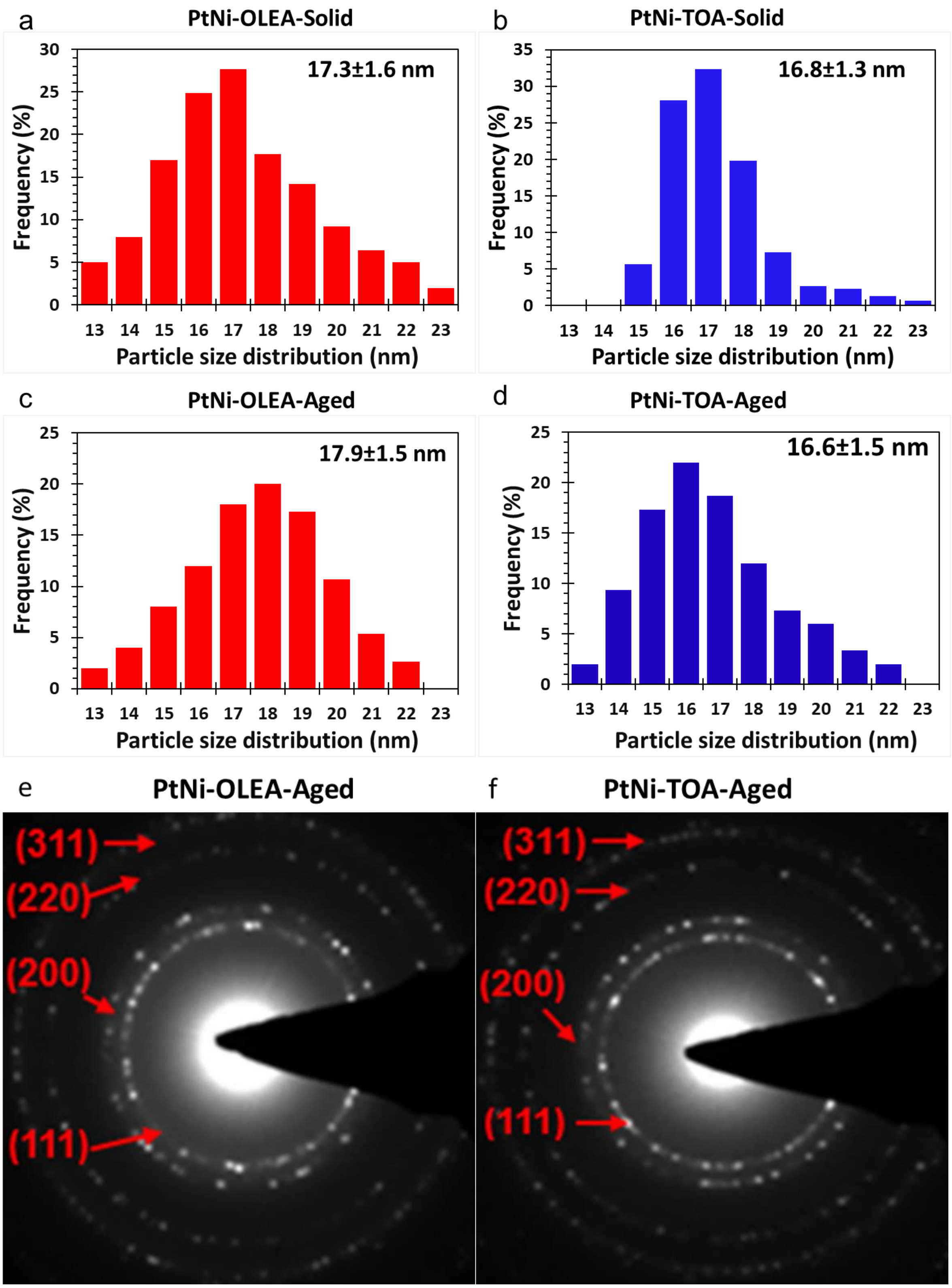

**Figure S1.** Nanostructures synthesized using high boiling point solvent 1-octadecene. (**a-d**) size distribution histograms of PtNi-OLEA-Solid, PtNi-TOA-Solid, PtNi-OLEA-Aged and PtNi-TOA-Aged nanoparticles, respectively. (**e,f**) Selected area electron diffraction patterns for both PtNi-OLEA-Aged and PtNi-TOA-Aged nanoparticles indicate the same dominant crystallographic features as those found in PXRD.

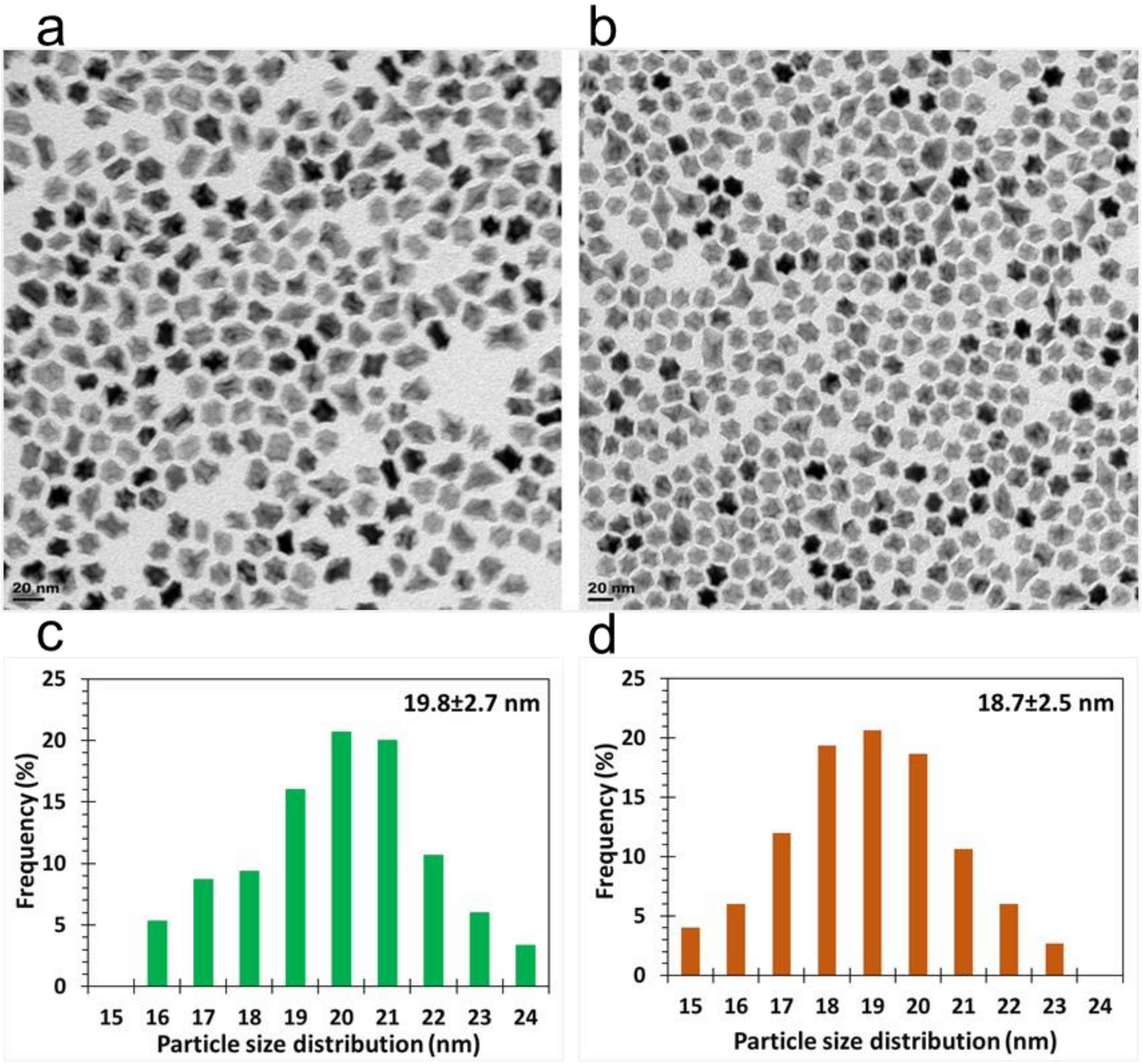

**Figure S2.** Nanostructures synthesized using low-boiling point solvent benzyl ether, showing excellent dispersity. (**a,b**) Overview TEM images of PtNi-OLEA-Aged and PtNi-TOA-Aged, respectively. (**c,d**) size distribution histograms of PtNi-OLEA-Aged and PtNi-TOA-Aged, respectively.

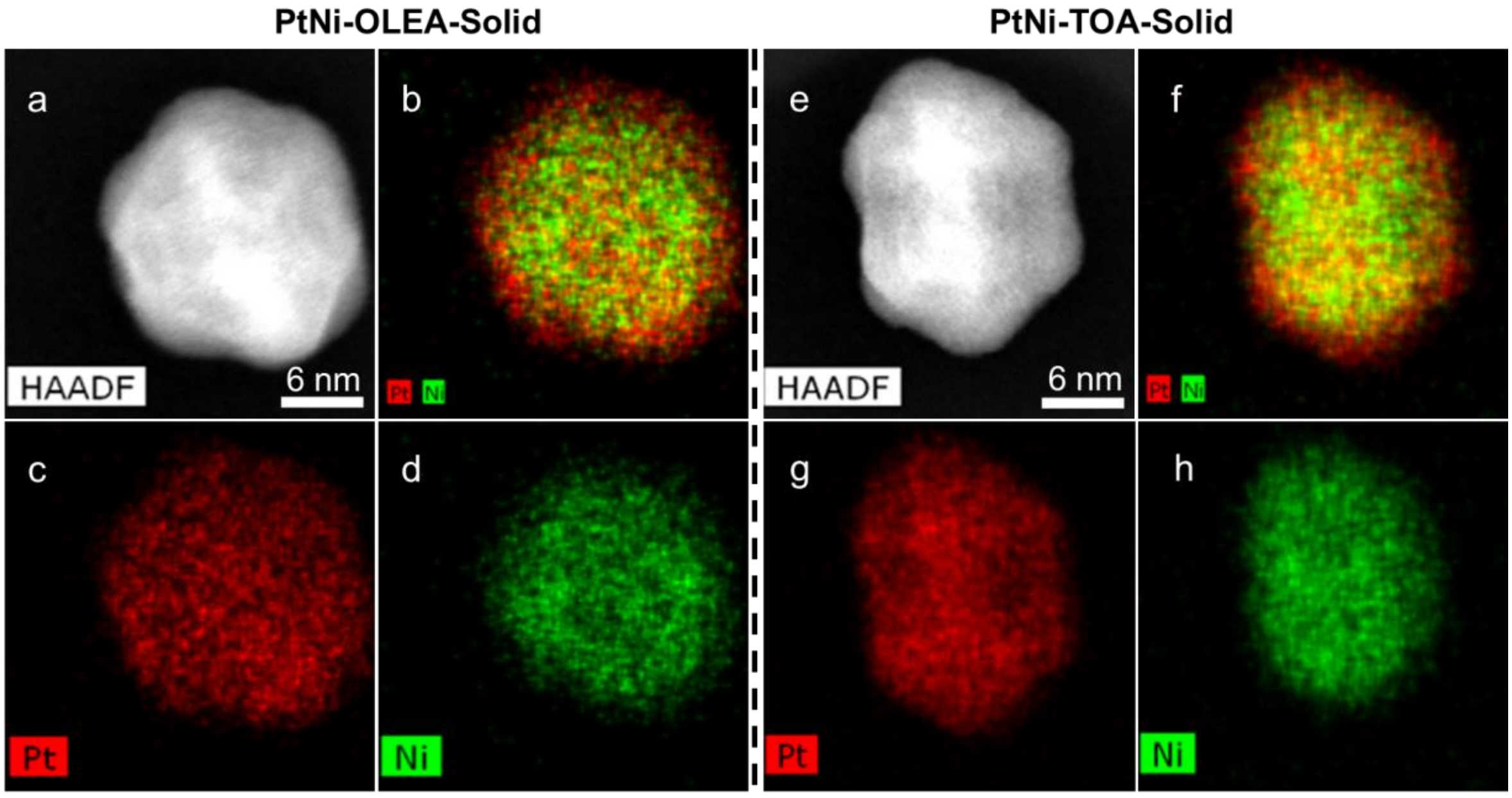

**Figure S3.** STEM-EDXS analysis of elemental distribution for **(a-d**) PtNi-OLEA-Solid nanoparticles and (**e-h**) PtNi-TOA-Solid nanoparticles. (**a,e**) show HAADF STEM images, (**b,f**) are EDXS composite maps of Pt and Ni, (**c,g**) are EDXS Pt maps and (**d,h**) EDXS Ni maps.

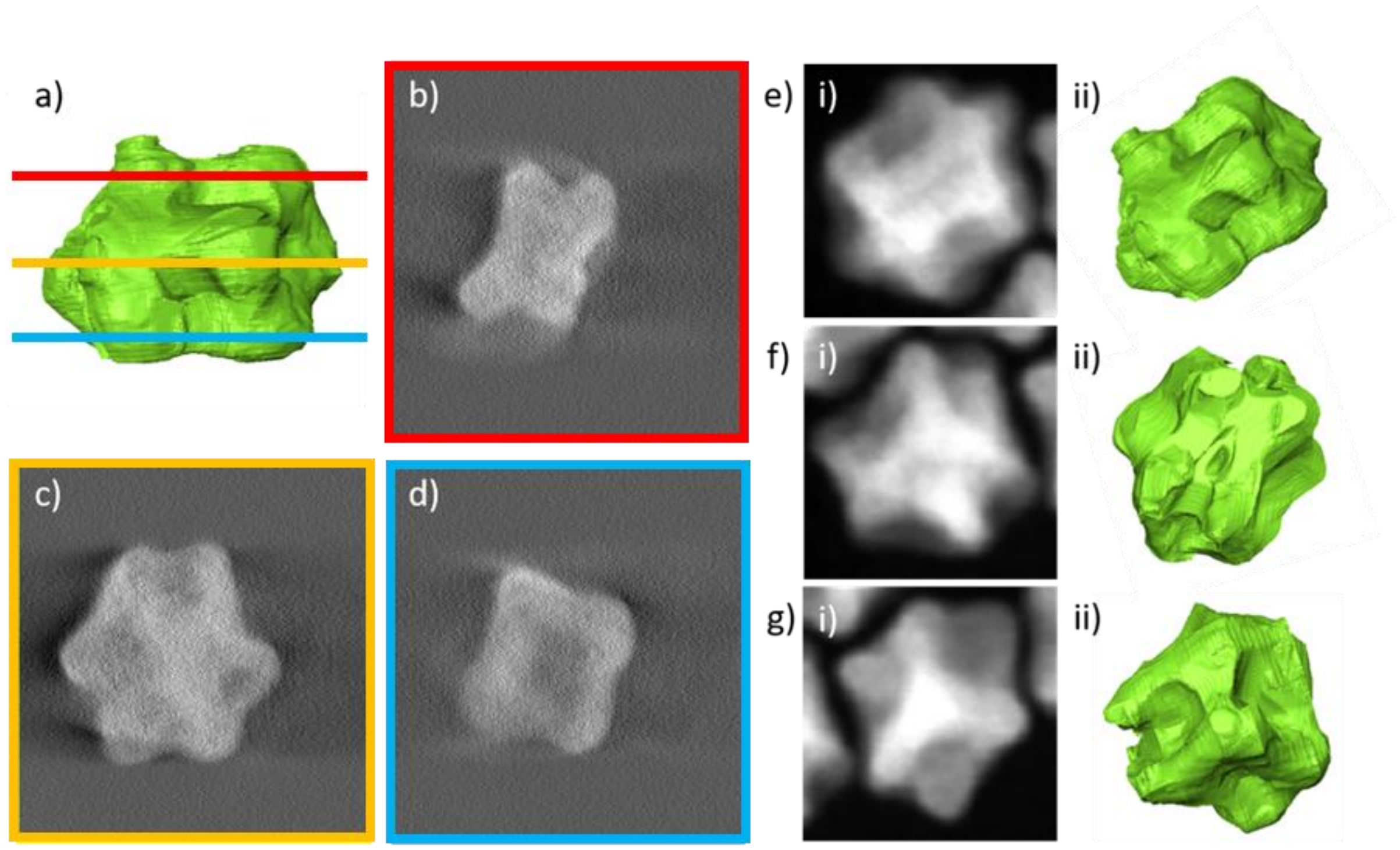

**Figure S4.** STEM HAADF tomography reconstruction of a PtNi-OLEA-Aged nanoparticle. (**a**) Surface visualisation of the nanoparticle and slices through the (**b**) top, (**c**) middle and (**d**) bottom of the nanoparticle, as indicated by the lines in (**a**). The reconstruction demonstrates a clear rhombic dodecahedron structure. (**e-f**) Comparison of HAADF STEM images (**i**) with surface visualisations (**ii**) of the tomographic reconstruction in a similar orientation.

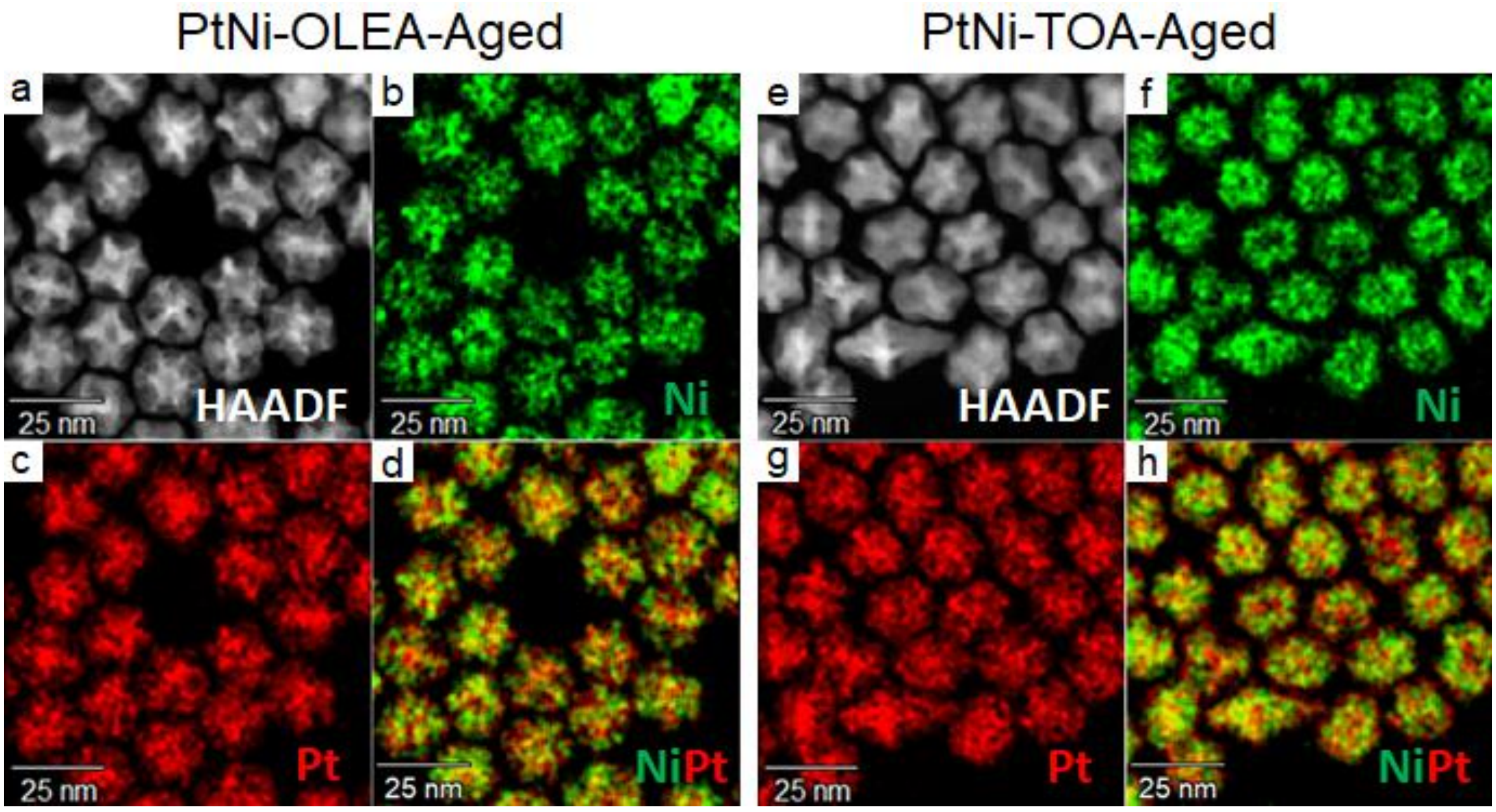

**Figure S5.** STEM-EDXS analysis of elemental distribution for **(a-d**) PtNi-OLEA-Aged nanoparticles and (**e-h**) PtNi-TOA-Aged nanoparticles. (**a,e**) show HAADF STEM images, (**b,f**) are EDXS Ni maps, (**c,g**) EDXS Pt maps and (**d,h**) EDXS composite maps of Pt and Ni.

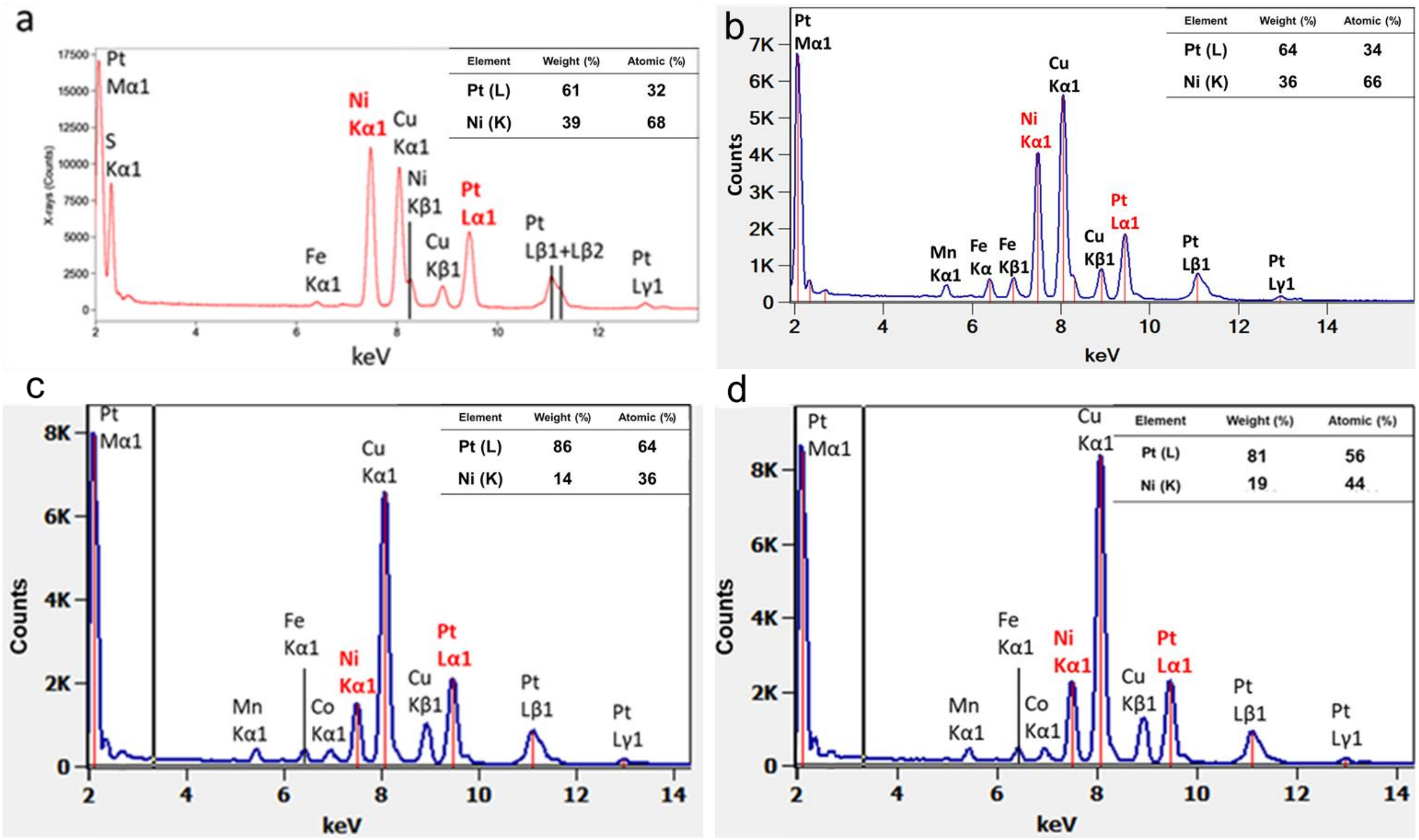

**Figure S6.** EDX spectra of (**a**) PtNi-OLEA-Solid, (**b**) PtNi-TOA-Solid, (**c**) PtNi-OLEA-Aged and (**d**) PtNi-TOA-Aged nanoparticles. All detected X-ray peaks in the energy range of 2 keV to 14 keV are labelled. Peaks in range of 0-2 keV are not shown due to the large number of overlapping elements present in this energy range. The Cu $K_{\alpha1}$ and Cu $K_{\beta1}$ peaks result from the use of a Cu TEM support grid, Mn $K_{\alpha1}$ from the use of a Mn washer to secure the grid. Peaks from Fe and Co are artefacts due to scattering from the polepiece. The presence of S $K_{\alpha1}$ in the PtNi-OLEA-Solid EDX spectrum is expected to be due to contamination. Major peaks used for quantification of the nanoparticle composition are labelled in red.

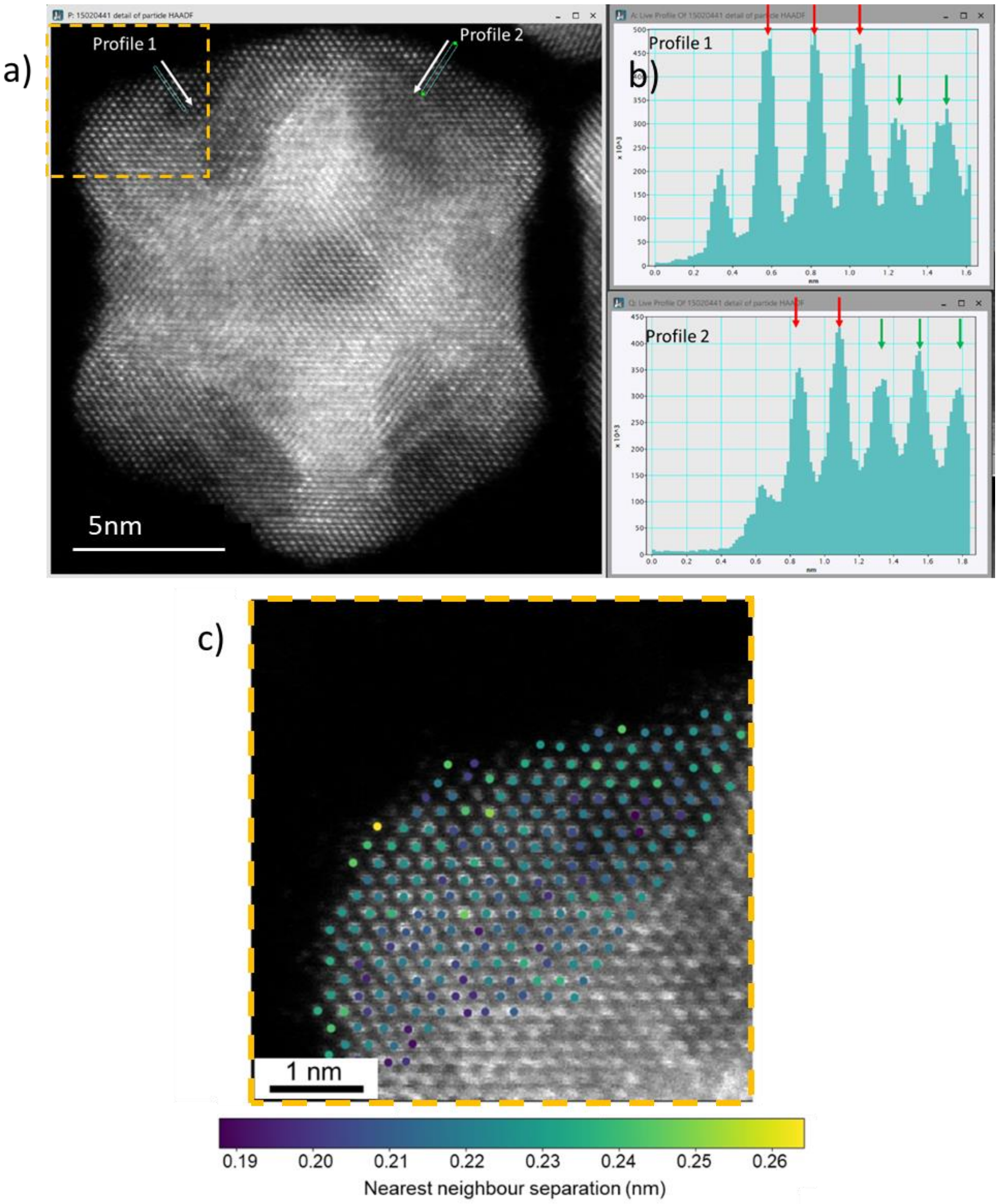

**Figure S7.** Atom column intensity and position analysis for a PtNi-TOA-Aged nanoparticle. a) HAADF STEM image, b) line profiles from the positions marked as white arrows and cyan rectangles in (a). On the profiles red arrows indicate atomic columns close to the surface of the sample where the HAADF intensity increases unexpectedly. For uniform composition the thickness is expected to decrease towards the edge of the particle, so this suggests these surface layers contain more of the higher atomic number species (Pt). Thus the Pt rich surface layer is estimated as ~3-4 atomic columns; consistent with the Pt enrichment observed in STEM-EDXS mapping. c) Analysis of projected nearest neighbour distance for a region of the particle shown by the dashed orange square in (a). Here we measure an average value of ~0.22 ±0.01 nm, consistent with the expected value for the {111} spacing in PtNi-TOA-Aged particles as measured via PXRD and electron diffraction (~0.220 nm see table S1). Qualitatively slightly larger interlayer spacings appear to be associated with the surface layer (3-4 atomic columns).

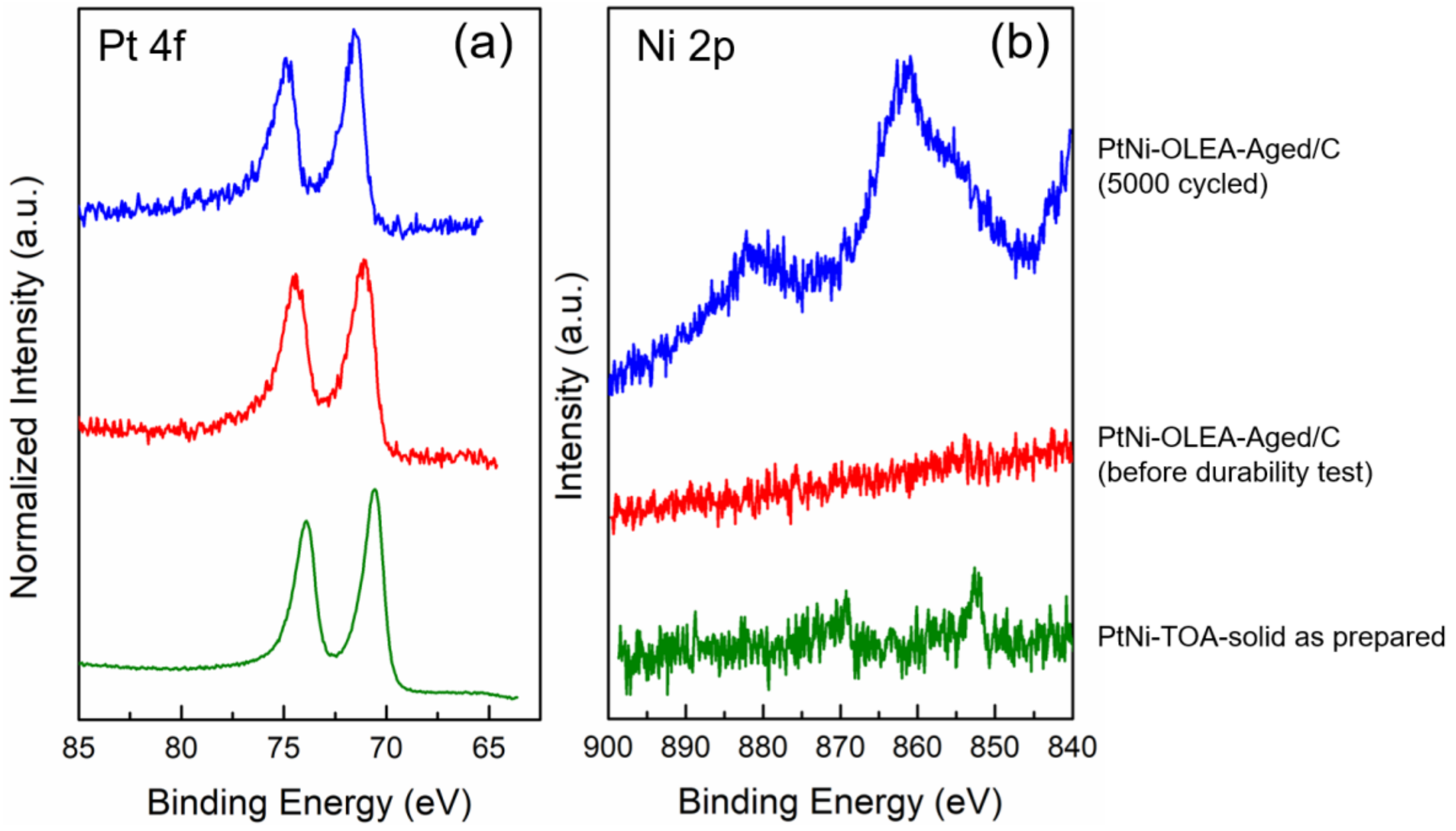

**Figure S8.** Normalised Pt 4f (**a**) and Ni 2p (**b**) high-resolution XPS spectra for PtNi nanoparticles: as-prepared, Aged/C and Aged/C after 5000 cycles. Results for TOA and OLEA routes are similar. That the XPS Ni peak is much weaker than the Pt peak in the as-prepared sample, which has an average composition of $Pt_1Ni_2$, is suggestive of a core/shell structure with an outer Pt rich shell. In this case, the shell thickness should be similar to the information depth of the XPS measurement, defined as 3* Inelastic Mean Free Path (IMFP) of electrons from the Ni 2p peak in Pt metal. The IMFP of 630 eV photoelectrons in Pt metal using the TPP formula[16] at 0.95 nm is calculated to be 2.85 nm, which indicates the lower limit for the Pt shell thickness is around 2.85 nm, consistent with the STEM-HAADF analysis of a shell thickness of 3-4 atomic layers (Figure 1(h-m)).

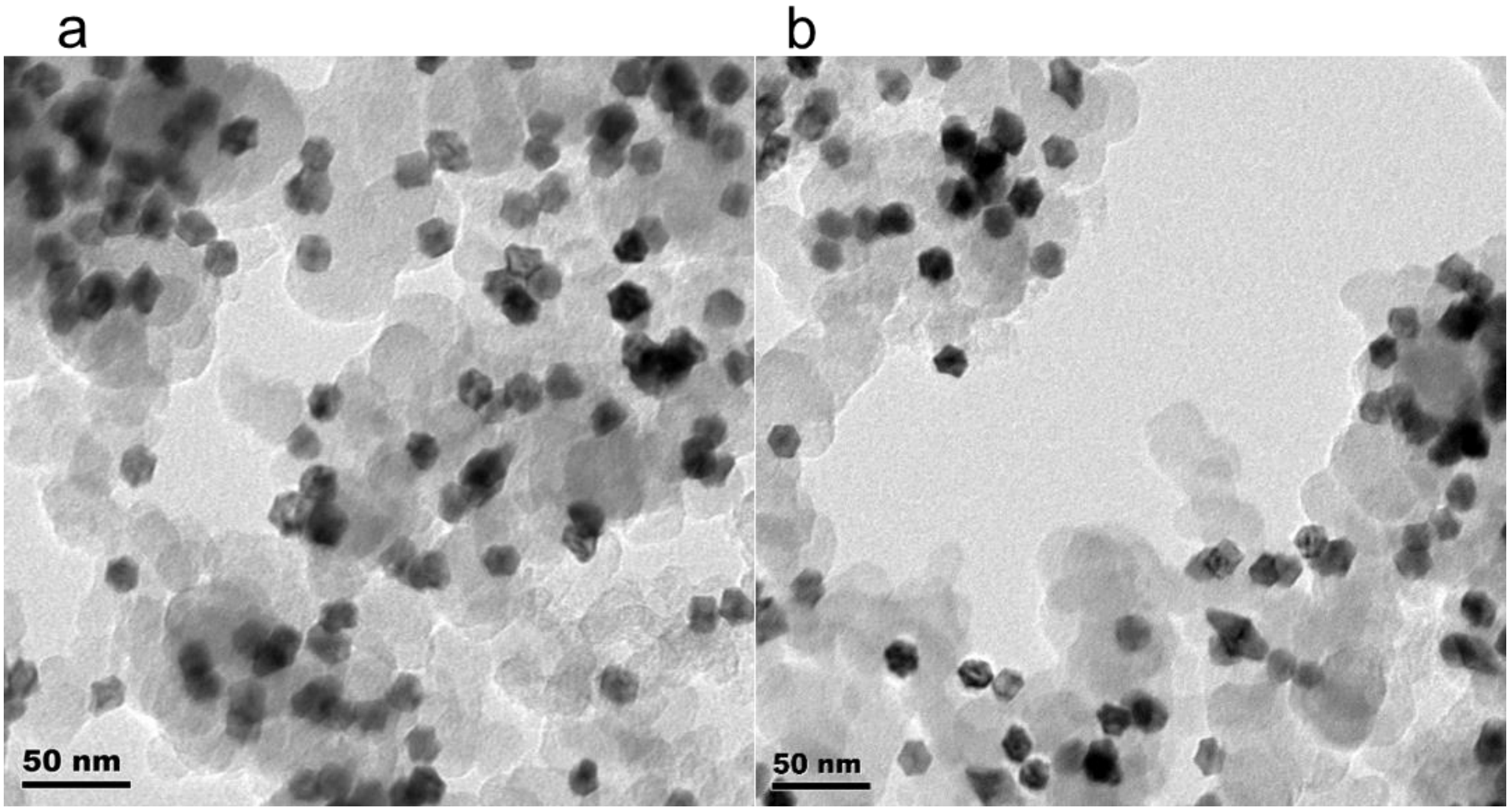

**Figure S9.** BF-TEM images of (**a**) PtNi-OLEA-Solid/C and (**b**) PtNi-TOA-Solid/C nanoparticles.

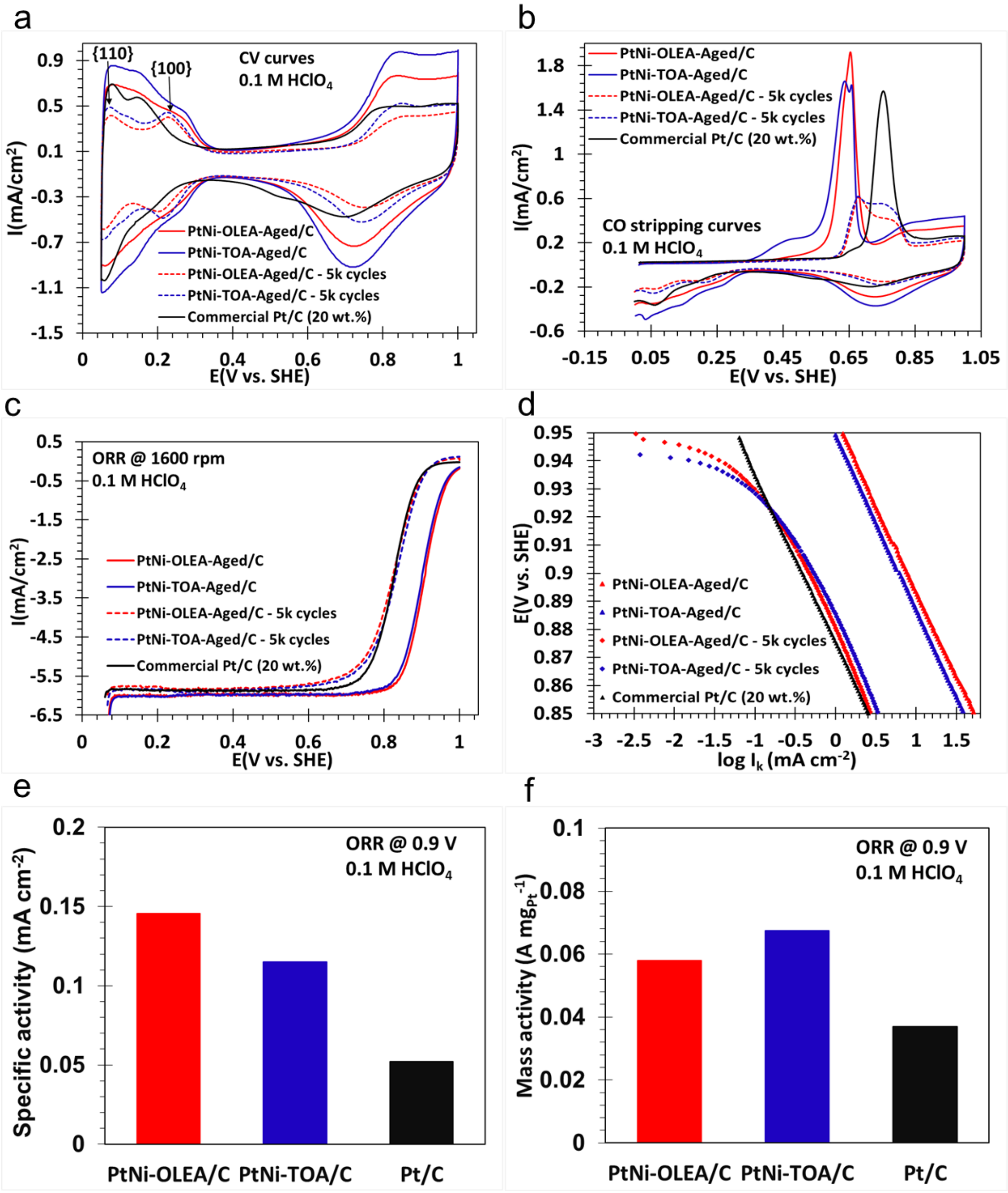

**Figure S10.** (**a**) Cyclic voltammograms of binary PtNi-OLEA-Aged/C (red), PtNi-TOA-Aged/C (blue) and commercial Pt/C (black) electrocatalysts showing the evolution of two peaks (indicated) indexed to {100} and {110} planes, (**b**) CO-stripping voltammetry curves, (**c**) ORR polarization curves, (**d**) the corresponding Tafel plots, (**e**) intrinsic area-specific activities and (**f**) mass-specific activities at +0.9 V (vs SHE), post 5000 electrochemical cycles.

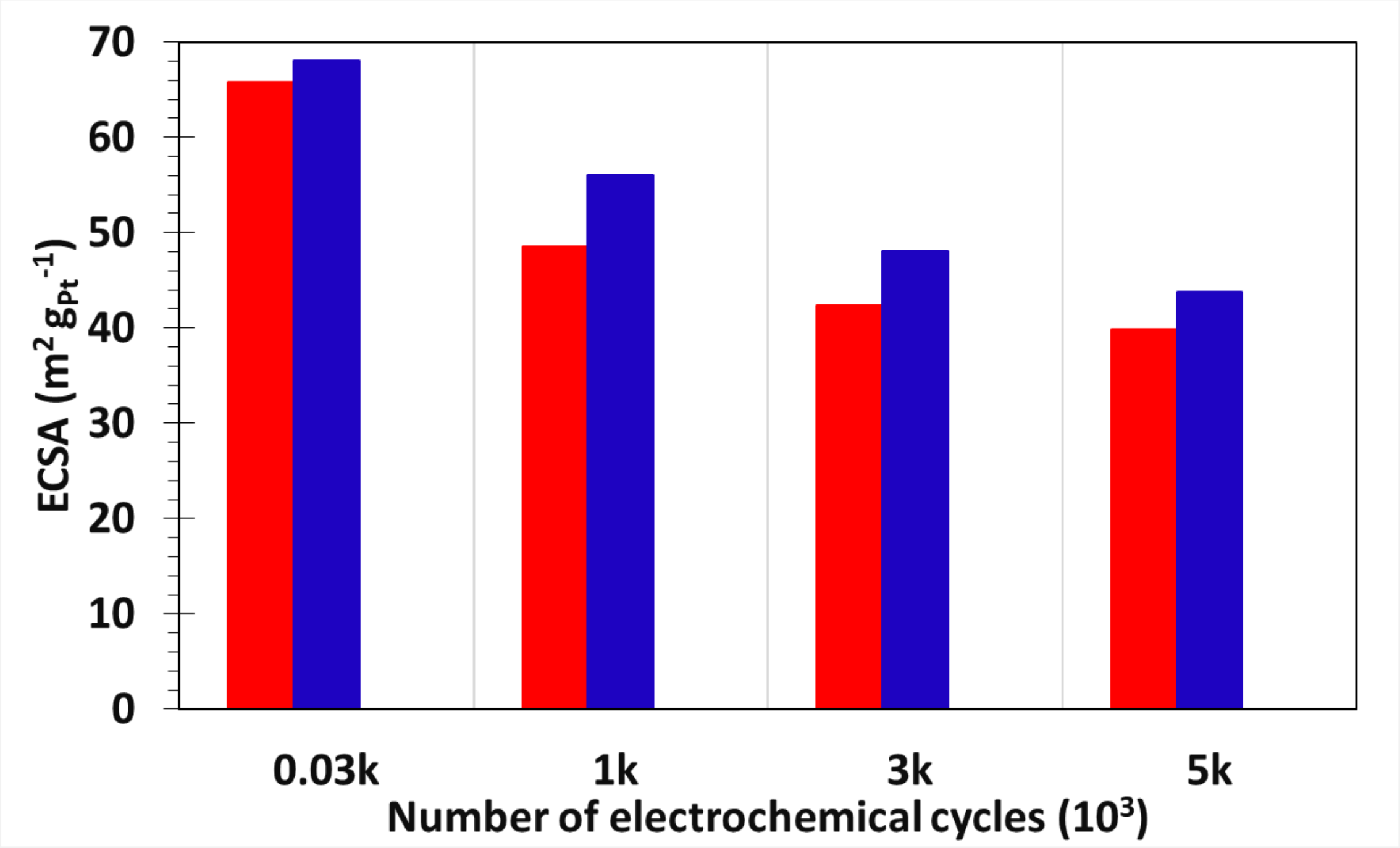

**Figure S11.** ECSA decay as a function of the continual potential cycling (up to 5000 cycles) of PtNi-OLEA-Aged/C (red) and PtNi-TOA-Aged/C (blue) binary nanoparticles.

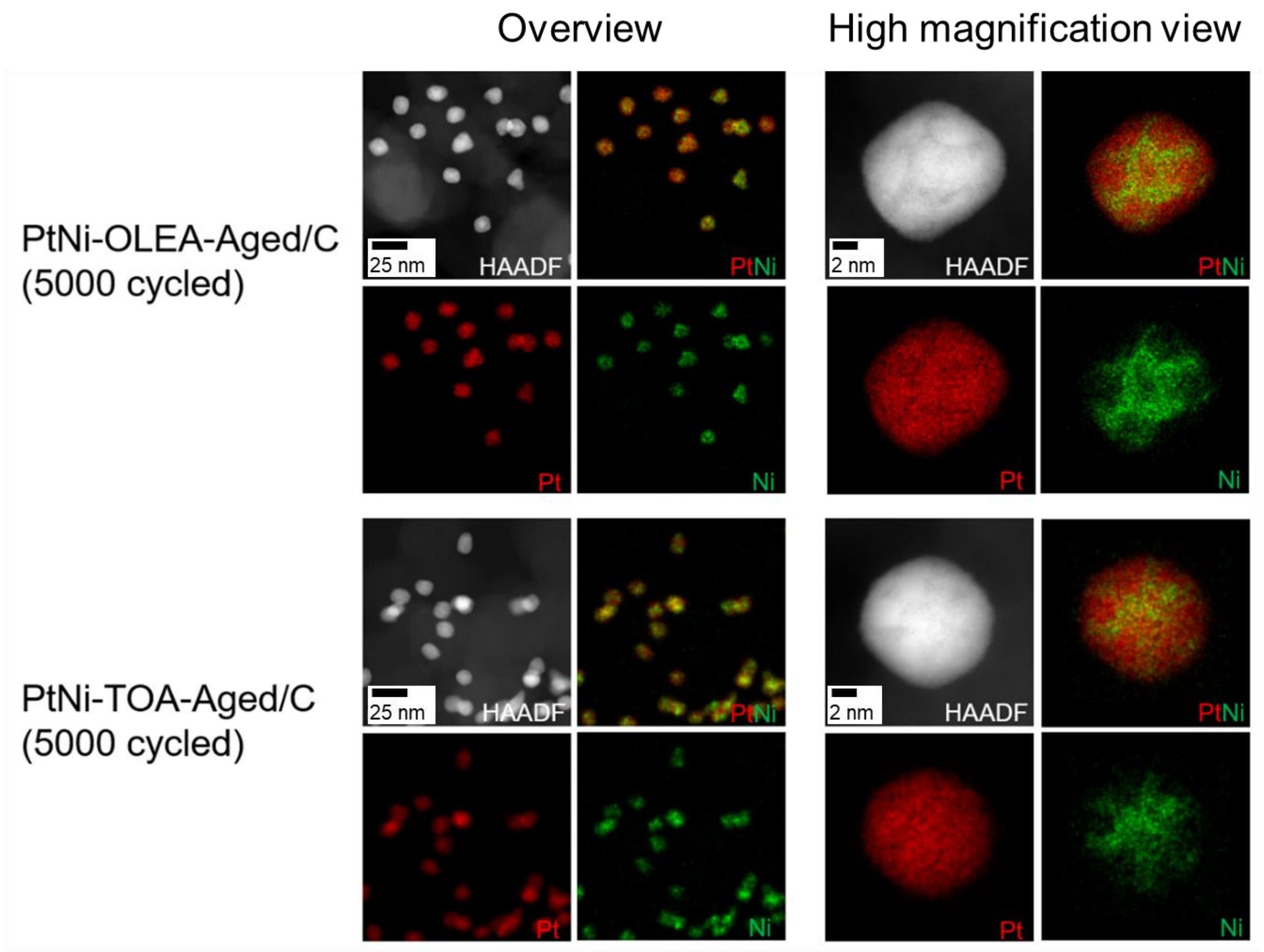

**Figure S12.** HAADF STEM and STEM EDXS elemental maps of PtNi-OLEA-Aged/C and PtNi-TOA-Aged/C nanoparticles after 5000 continual potential cycling.

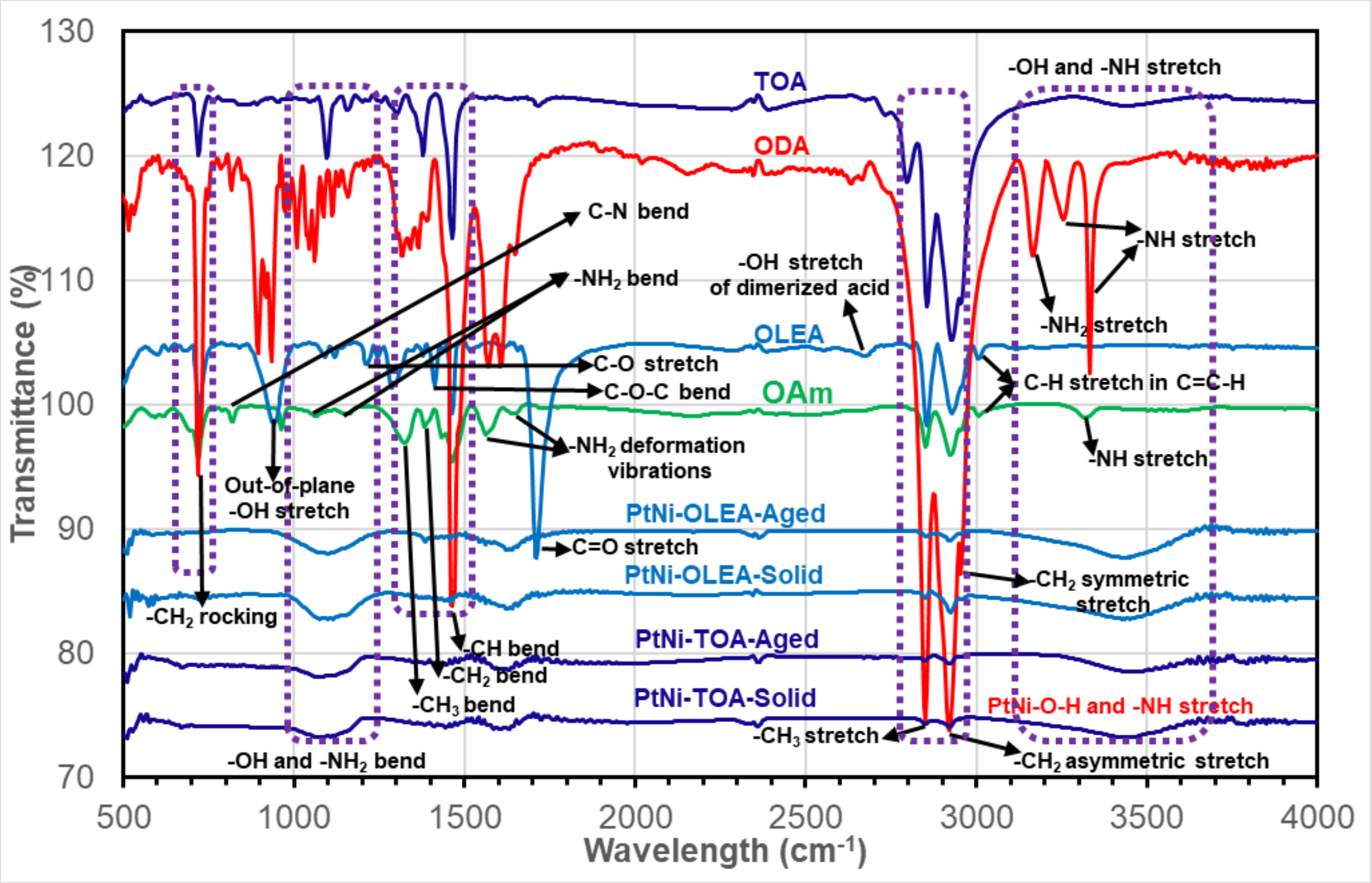

**Figure S13**. Fourier transform infrared measurements of all surfactants (TOA, ODA, OLEA, OAm) and all samples (PtNi-OLEA-Solid, PtNi-OLEA-Aged, PtNi-TOA-Solid, PtNi-TOA-Aged).

**Table S1.** Measured and calculated d-spacings and lattice parameters. Note that for bulk fcc Pt and Ni the lattice constants are 0.3912 nm and 0.3499 nm,[17] respectively.

| Sample | XRD | | | HAADF-STEM | | TEM Diffraction | |
|---|---|---|---|---|---|---|---|
| | 2θ diffraction peaks (°) | Measured d-spacing (nm) | Calculated lattice parameter (nm) | Measured d-spacing (nm) | Calculated lattice parameter (nm) | Measured d-spacing (nm) | Calculated lattice parameter (nm) |
| PtNi-OLEA-Solid | 41.534 | 0.217 {111} | 0.376 | 0.22 ± 0.01 {111} | 0.38 ± 0.02 | - | - |
| PtNi-OLEA-Aged | 40.693 | 0.222 {111} | 0.384 | 0.19 ± 0.01 {200} | 0.38 ± 0.02 | - | - |
| PtNi-TOA-Solid | 41.841 | 0.216 {111} | 0.374 | 0.22 ± 0.01 {111} | 0.38 ± 0.02 | 0.215 ± 0.005 {111}<br>0.185 ± 0.005 {200} | 0.372 ± 0.009<br>0.370 ± 0.010 |
| PtNi-TOA-Aged | 41.023 | 0.220 {111} | 0.381 | 0.22 ± 0.01 {111} | 0.38 ± 0.02 | 0.222 ± 0.005 {111}<br>0.190 ± 0.005 {200} | 0.385 ± 0.009<br>0.381 ± 0.010 |

**Table S2.** Comparisons of the $ECSA_{CO}$, $ECSA_{Hupd}$, mass-specific and area-specific activities at 0.9 V and $ECSA_{CO}/ECSA_{Hupd}$ of two binary electrocatalysts before and after extended durability measurements.

| Electrocatalysts | ECSA ($m^2/g_{Pt}$) (CV) | ECSA ($m^2/g_{Pt}$) (CV) loss (%) | ECSA ($m^2/g_{Pt}$) (CO stripping) | ECSA ($m^2/g_{Pt}$) (CO stripping) loss (%) | Mass activity ($A/mg_{Pt}$) | Mass activity loss (%) | Specific activity ($mA/cm^2_{Pt}$) | Specific activity loss (%) |
|---|---|---|---|---|---|---|---|---|
| PtNi-OLEA-Solid/C | 0.1k cycles: 56.2 | - | - | - | 0.33 | - | 0.59 | - |
| PtNi-TOA-Solid/C | 0.1k cycles: 59.4 | - | - | - | 0.30 | - | 0.51 | - |
| PtNi-OLEA-Aged/C | 0.03k cycles: 65.7 | 39.4 | 0.03k cycles: 71.7 | 39.5 | 0.91 | 93.7 | 1.39 | 90 |
| | 5k cycles: 39.8 | | 5k cycles: 43.4 | | 0.06 | | 0.15 | |
| PtNi-TOA-Aged/C | 0.03k cycles: 68.1 | 35.7 | 0.03k cycles: 74.7 | 32.9 | 0.75 | 91.0 | 1.10 | 90 |
| | 5k cycles: 43.8 | | 5k cycles: 50.1 | | 0.07 | | 0.12 | |
| Pt/C | 0.1k cycles: 71.0 | - | 0.1k cycles: 87.2 | - | 0.037 | - | 0.052 | - |

| ECSA ($CO_{ad}$)/ECSA ($H_{upd}$) | PtNi-OLEA-Aged/C | PtNi-TOA-Aged/C |
|---|---|---|
| 0.03k cycles | 1.10 | 1.10 |
| 5k cycles | 1.10 | 1.14 |

**Table S3.** Comparison of the ORR activities over Pt alloy catalysts in recent studies

| Catalyst | Particle size (nm) | Metal loading | Electrochemical surface area ($m^2 g_{Pt}^{-1}$) | Area-specific activities ($mA \cdot cm^{-2}$) | Mass-specific activities ($A \cdot mg_{Pt}^{-1}$) |
|---|---|---|---|---|---|
| $Pt_{1.5}Ni$ octahedral/C[18] | 5.8±1.5 | 20.4wt% | 48.3 $^{a}$ | 3.99$^{a}$ | 1.96 |
| PtNi octahedra/C[19] | 7.98±0.38 | 29.4wt% | 45.0 | / | ~1.8 |
| PtNi octahedra/C[20] | 9 | ~25wt% | 45.0 $^{b}$ | 10.1$^{b}$ | 3.30 |
| PtNi octahedral/C[21] | 9.5±0.8 | / | 50.0 $^{b}$ | 3.14$^{b}$ | 1.45 |
| $Pt_3Ni$ octahedra/C[22] | 10 | 20wt% | / | 1.26 | 0.44 |
| $Pt_2Ni$ octahedra/C[23] | 9.5±0.5 | / | / | / | 2.0 |
| PtNi octahedra/C[24] | 12±0.8 | / | / | ~3.8$^{b}$ | 1.70 |
| Nanoporous NiPt/C[25] | 15 | / | 41.0 $^{a}$ | 3.34$^{a}$ | 1.32 |
| PtNi cubes/C[26] | 8-9 | / | / | 3.0$^{a}$ | 0.68 |
| $Pt_3Ni$ octahedra/C[27] | ~4 | / | 66.6 $^{a}$ | 2.7$^{a}$ | 1.80 |
| $Pt_3Ni$ octahedra/C[28] | 11 | / | / | 2.8$^{a}$ | 0.11 |
| PtNi octahedra/C[29] | 7.10 ± 1.1 | / | 38.1 $^{a}$ | 2.42$^{a}$ | 0.92 |
| PtNi hexapods/C[30] | 10-20 | / | 47.2 | / | 0.85 |
| $Pt_3Ni$ nanoframe/C[31] | ~20 | ~20wt% | / | 1.52$^{b*}$ | 5.7 |
| Jagged PtNi nanowires/C[32] | / | / | 118.0 $^{a}$ | 11.5$^{a}$ | 13.6 |
| $PtCo_3$ nanoparticles/C[33] | 8.9 ± 2.0 | / | 60.0 | 2.24 | 1.14 |
| PtCo nanocubes/graphene[12] | 5.5 ± 0.5 | / | 61.3 | ~0.5 | 0.95 |
| $Pt_{68}Cu_{32}$ nanoparticles[34] | 4.84± 0.07 | / | 12.1 $^{a}$ | 2.85$^{a}$ | 0.34 |
| PtCu nanoframes/C[35] | ~40 | ~20wt% | 66.2 $^{a}$ | 1.24$^{a}$ | 0.82 |
| **Pt/Ni-OLEA-Aged/C (This work)** | **17.9±1.5** | **~20wt%** | **65.7** $^{a}$ | **1.39**$^{a}$ | **0.91** |
| **Pt/Ni-TOA-Aged/C (This work)** | **16.6±1.5** | **~20wt%** | **68.1** $^{a}$ | **1.10**$^{a}$ | **0.75** |

Both area and mass specific activities are measured at 0.90 V in 0.1M $HClO_4$ electrolyte.
$^{a}$Normalized to the surface areas derived from the H adsorption charges. $^{b}$Normalized to the surface areas derived from the CO stripping charges. *Activities are measured at 0.95 V in 0.1M $HClO_4$ electrolyte.